\documentclass[a4paper, 12pt]{article}
\usepackage[margin=0.7in]{geometry}  
\usepackage{lineno} 

\usepackage{graphicx} 
\usepackage{amsfonts}       
\usepackage{breqn}
\usepackage{url}            
\usepackage{listings}
\usepackage{xcolor}
\usepackage{caption}
\usepackage{subcaption}
\usepackage{multirow} 
\usepackage{cite}
\usepackage[colorlinks=true, linkcolor=blue, citecolor=blue, urlcolor=blue]{hyperref}
\usepackage{float}
\usepackage[stable]{footmisc}
\usepackage{tablefootnote}
\usepackage{amsmath}
\usepackage{array}
\usepackage{setspace}
\usepackage{booktabs}
\usepackage{fontawesome5}  

\lstdefinestyle{pythonstyle}{
    language=Python,
    basicstyle=\ttfamily\small,
    breaklines=true,
    commentstyle=\color{green!60!black},
    keywordstyle=\color{blue},
    stringstyle=\color{orange},
    numbers=left,
    numberstyle=\tiny\color{gray},
    numbersep=10pt,
    tabsize=4,
    showstringspaces=false,
    frame=single,
    captionpos=b
}

\usepackage{tcolorbox}
\newtcolorbox{jupyteroutput}{
    colback=white,
    colframe=gray!20,
    boxrule=1pt,
    arc=0pt,
    outer arc=0pt,
    leftrule=2mm,
    rightrule=2mm,
    toprule=2mm,
    bottomrule=2mm,
    boxsep=0pt,
    left=2mm,
    right=2mm,
    top=2mm,
    bottom=2mm
}

\newcommand{\beq}{\begin{eqnarray}}
\newcommand{\eeq}{\end{eqnarray}}
\newcommand{\bpmatrix}{\begin{pmatrix}}
\newcommand{\epmatrix}{\end{pmatrix}}
\newcommand{\ba}{\begin{array}}
\newcommand{\ea}{\end{array}}

\def\bea{\begin{eqnarray}}   
\def\eea{\end{eqnarray}}

\title{Fast multilabel classification of HEP constraints with deep learning}

\author{
  Maien Binjonaid \\
  Department of Physics and Astronomy\\
  King Saud University\\
  Riyadh, Saudi Arabia \\
  \texttt{maien@ksu.edu.sa} \\
}

\begin{document}

\maketitle


\begin{abstract}
The shortcomings of the Standard Model (SM) motivate its extension to accommodate new expected phenomena, such as dark matter and neutrino masses. However, such extensions are generally more complex due to the presence of a large number of free parameters and additional phenomenology. Understanding how theoretical and experimental limits affect the parameter spaces of new models, individually and collectively, is of utmost importance for conducting model status analysis, motivating precise computations, or model-building aimed at solving certain issues. However, checking the constraints usually require a large amount of time using a chain of physics tools. We demonstrate, for the first time, the application of deep learning (DL) for the multilabel classification (MLC) of a group of theoretical and experimental constraints in the dark doublet phase of the next-to two-Higgs-doublet model (DDP-N2HDM), as a representative 9-dimensional parameter space. We analyze the issue of class imbalance and the ability of the classifier to learn joint class distributions. We demonstrate the time advantage compared to physics tools, with the classifier achieving orders of magnitude faster checks on groups of constraints and strong performance. The classifier performed strongly in terms of identifying regions where all constraints are valid or invalid, as well as regions where one or more of the constraints are valid or invalid simultaneously. This approach can be applied to any extension beyond the SM with the potential to aid HEP tools or act as a surrogate for fast model status checks. To that end, we provide a python tool \texttt{HEPMLC} for generating and investigating multilabel classifiers for SM extensions. 
\end{abstract}

\section{Introduction} \label{intro}
Machine learning (ML) and deep learning (DL), which are specific types of artificial intelligence (AI), have proven to be valuable tools in various research fields, including particle physics. These methods are particularly effective in tasks such as pattern recognition, anomaly detection, classification, and regression problems, often offering greater efficiency than traditional techniques ~\cite{GoodBengCour16, 10.5555/2371238}. AI has a wide range of applications in both theoretical and experimental studies of the Standard Model (SM) and its extensions, collectively called Beyond the SM (BSM). These can be grouped into several areas, including signal-to-background analysis, jet-tagging, parameter space scanning, learning properties and constraints, and predicting likelihoods, among many other applications (for example \cite{Guest:2018yhq, Albertsson:2018maf, Radovic:2018dip, Bourilkov:2019yoi, Carleo:2019ptp, Abdughani:2019wuv, Qu:2019gqs, Moreno:2019neq, Bogatskiy:2020tje, Schwartz:2021ftp, Karagiorgi:2022qnh, Qu:2022mxj, Shanahan:2022ifi, Alda:2021rgt, ParticleDataGroup:2024cfk}). A comprehensive review that is constantly updated can be found in Ref.~\cite{Feickert:2021ajf}.

As is well known, BSM extensions aim to address issues not explained by the SM, such as the origin of neutrino masses, the strong CP problem, the existence of Dark Matter (DM), which may well be of particle nature, and the lightness of the SM Higgs particle \cite{Gaillard:1998ui}. Models in this context often involve complex, multidimensional parameter spaces with numerous free parameters and new particles. In utilizing AI, some studies have focused on BSM signal-to-background analysis \cite{Cogollo:2020afo, Hammad:2022lzo, Esmail:2023axd, Belfkir:2023vpo}, flavor phenomenology \cite{Alda:2021rgt}, Higgsino jets \cite{PhysRevD.106.055008}, and R-parity violating supersymmetry \cite{PhysRevD.98.076017}.

The applications of AI in the process of scanning and analyzing constraints is progressing, but many aspects are still at an exploration phase as lessons are being learned about different methods. As for analyzing and classifying the properties of parameter spaces using AI, Ref.~\cite{Caron:2016hib} presented one of the earliest uses of AI to classify parameter points in the Minimal Supersymmetric Model (MSSM) as valid or invalid, based on ATLAS data, using a Random Forest (RF) classifier. This was followed by an Active Learning network \cite{Caron:2019xkx} that consistently outperformed RF classifiers and significantly improved scanning times compared with random sampling. In Ref.~\cite{Wang:2020tap}, a Heuristic Search was combined with a Generative Adversarial Network to classify the validity of parameter points in the Next-to-Minimal Supersymmetric SM (NMSSM). A Generative Normalization Flow method was used in Ref.~\cite{Hollingsworth:2021sii} to explore the MSSM, only taking into account the mass of the SM Higgs boson as a constraint.

In Ref.~\cite{Baruah:2024gwy}, a trained neural network was proposed to predict higher likelihood points for a nested sampling algorithm, and applied to a toy model. Ref.~\cite{Ren:2017ymm} introduced a criterion for scanning parameter spaces using active learning, which was implemented in Ref.~\cite{Staub:2019xhl}. A \texttt{Mathematica} tool was presented in Ref.~\cite{Arroyo-Urena:2020qup} to set limits on the free parameters of a BSM models, and produce benchmark points using ML. Moreover, the work presented in Ref.~\cite{deSouza:2022uhk} introduced a sampling technique that utilizes ML Black-Box optimization and applied it to the MSSM, potentially improving scanning efficiency by orders of magnitude, although only the mass of the SM Higgs boson was taken as a constraint or in combination with the constraint from dark matter relic density. A scanning tool was introduced in \cite{Goodsell:2023iac} which enables active learning scans within the public package \texttt{SARAH} \cite{Staub:2013tta}. 

However, most classification studies in the literature focus on single-label classification (SLC), where the target output is a binary class (i.e. valid or invalid). Specifically, when applying AI in exploring parameter spaces of BSM extensions, using specific physics tools/packages, to validate inputs against constraints, a single-label classifier would learn if the set of input parameters lead to an overall-valid/invalid result. That is, the physics tool determined that all constraints were passed (valid result) or that at least one constraint is not satisfied (invalid result). Therefore, the resulting trained AI classifier has no information regarding the status of the underlying constraints (they were hidden from the classifier as the decision of overall validity/invalidity was done by the physics tool and only the final decision is passed to the AI single-label classifier). While this approach has its own merit, and can be informative and efficient in aiding parameter space scans targeting only valid points, the nature of SLC is clearly restrictive since it cannot learn and provide insights into how each constraint (or category of constraints) affects the outcome or how different constraints might interact. 

Understanding these aspects, regardless of AI, has been an integral part of particle phenomenology in studies that aim to systematically analyze the effects of constraints on a given model (i.e. status of the model), motivate or investigate effects of precise computations (e.g. higher-order corrections) or extensions that would open or close certain regions (See for e.g. \cite{Baer:2002gm, Ellis:2002wv, Domingo:2007dx, Djouadi:2008uw, Djouadi:2008yj, Buchmueller:2012hv, Celis:2013rcs, Eberhardt:2013uba, Craig:2013cxa,  Buchmueller:2014yva, deVries:2015hva, Staub:2015kfa, Staub:2017ktc, GAMBIT:2017gge, Krauss:2017xpj, Goodsell:2018tti, Krauss:2018orw, Goodsell:2018fex, Staub:2018vux, Baer:2020kwz, Ellis:2022emx, Treesukrat:2024akz, King:1995ys , Masip:1998jc, Barbieri:2007tu, King:2012is, LopezHonorez:2006gr, Barbieri:2006dq, Arhrib:2013ela, LopezHonorez:2010tb, Arroyo-Urena:2019zah, Dhargyal:2017cqo}). These constraints can be theoretical or experimental. A well known set of theoretical constraints are vacuum stability, triviality and unitarity, which were essential in pinpointing the allowed range of the SM Higgs mass prior to its discovery \cite{Gunion:1989we}. Such constraints become more complicated in extensions of the SM, and are essential to take into account. Experimental constraints include electroweak precision tests, Higgs properties (e.g. couplings to SM particles), DM observations (if the model include a DM candidate), and flavor physics (e.g. B physics). All of these constraints and others that might be model-specific (e.g. related to supersymmetry) must be considered. In BSM extensions, the relevant computations are based on the set of input parameters of a given model, and involves sophisticated calculations of every theoretical and observable aspect. This prolonged process has been advanced by a number of numerical tools and packages\footnote{An extensive review of model building and the issues surrounding it is provided in \cite{ Ivanov:2017dad}.}. One example, which we utilize in this work, is \texttt{ScannerS} \cite{Muhlleitner:2020wwk}, enabling phenomenological analyses of extensions of the SM with doublets and singlets. The package includes routines to check theoretical and experimental constraints, and interfaces with \texttt{N2HDECAY} \cite{Engeln:2018mbg} (Higgs properties), \texttt{micrOMEGAs} \cite{Belanger:2020gnr} (DM observables), \texttt{HiggsBounds} \cite{Bechtle:2020pkv} and \texttt{HiggsSignals} \cite{Bechtle:2020uwn} (Higgs constraints), \texttt{EVADE} \cite{evade}  (vacuum structure).

Consequently, when it comes to the potential application of AI for detailed analyses of constraints in extensions of the SM, and given that SLC is not suitable for this task, we are led to Multilabel Classification (MLC) \cite{tsoumakas2007multi, deCarvalho2009, Herrera2016, TAREKEGN2021107965}, which has not been explored yet in this context. MLC addresses problems in which a set of inputs (features) is associated with multiple outputs (target labels), each of which can be binary or multiclass. Applying MLC to the study of parameter constraints, means that the AI classifier learns the decision of the physics tools on the validity of each constraint (or category of constraints) individually, but it will also implicitly learn each combination of constraints (or category of constraints) collectively (i.e. the joint distribution of constraints/labels). So while the physics tools are concerned with applying individual constraints given a set of inputs, the classifier is learning patterns between the inputs associated with the joint distribution of constraints. This means that once a MLC is trained on a BSM extension, and depending on the required details and performance, it can potentially play the role of the chain of physics tools in validating the model and providing the insights that are usually provided by such tools (i.e. acting as a surrogate). The potential of such method can be judged by achieving reasonable accuracy and time advantage.

MLC is significantly more involved than binary SLC, and it brings about new challenges such as dealing with class imbalance between combinations of labels. Hence, it is essential to test and validate its application within the context of a realistic BSM parameter space. It is then the goal of this study to develop a DL model for the MLC of parameter space constraints, focusing on learning their validity and invalidity, and demonstrating the feasibility of this method. We select four categories of constraints representing theoretical and experimental limits. Namely: Boundedness-from-blow (BFB), Perturbative Unitatiry (PU), Oblique parameters (STU), and bounds on SM-like and non-SM Higgs (Higgs). The classifier should be able to correctly predict the outcomes of individual constraints and any possible combinations. The success of the method is to be determined by appropriate performance-measuring metrics. The major one being the subset accuracy, which is a strict metric that considers an instance as correctly classified only if all labels are correctly identified at the same time.

To apply this method, we consider, as a representative case, the Dark Doublet Phase of the Next-to Two-Higgs Doublet Model (DDP-N2HDM). The model consists of 9 free parameters, creating a multidimensional parameter space that can be challenging to analyze, making it a suitable case for applying AI techniques. Moreover, the model consists of two Higgs doublets and a real singlet. One doublet remain inert (i.e. does not acquire a Vacuum Expectation Value (VEV)), thereby providing a candidate for DM that is stabilized through an unbroken discrete symmetry. The particle content of this model comprises two CP-even Higgs bosons (one of which is SM-like), and a dark sector containing a CP-even, CP-odd, and charged scalars, where the lightest neutral scalar represents the DM candidate. This model is interesting since its DM sector resembles that of the the Inert Doublet Model \cite{Branco:2011iw, Belyaev:2016lok, Engeln:2020fld}. While its Higgs sector resembles that of Type 1 2HDM. However, the presence of a singlet component provides a distinctive feature to its Higgs bosons. The model has a rich phenomenology, which has been studied in several works (see, for e.g., \cite{PhysRevD.89.075009, Drozd:2014yla, Muhlleitner:2016mzt, Ferreira:2019iqb, Azevedo:2021ylf, Biekotter:2021ysx, Glaus:2022rdc, Binjonaid_2024} and references therein).  

To facilitate future work on other models and studies dedicated to further advance this application (e.g. optimization studies for MLC, generate multilabel classifiers for commonly studied models), we provide a python tool \texttt{HEPMLC} that enables users to build multilabel classifiers using ML and DL for models within the public packgae $\texttt{ScannerS}$ as a first stage, or for any labeled dataset provided by the user. 

The paper is divided as follows: Section 2 introduces the N2HDM and details the constraints considered in this study. Section 3 provides a brief overview of MLC. Section 4 covers the data generation using the physics tool, the AI methodology, and the DL architecture. Section 5 presents the results, followed by discussion and conclusions in Section 6. Finally, we describe the python tool \texttt{HEPMLC} in the Appendix.

\section{The Physics Model and Constraints}
\label{sec:core1}

\subsection{The N2HDM}
\label{core:model}
We start with a brief overview of the N2HDM. The scalar potential is given by (following the notation in Ref.~\cite{Engeln:2020fld})

\begin{align}
V =&\enspace m_{11}^{2} \Phi_{1}^{\dagger} \Phi_{1} + m_{22}^{2} \Phi_{2}^{\dagger} \Phi_{2}
+ \dfrac{\lambda_{1}}{2} \left(\Phi_{1}^{\dagger} \Phi_{1}\right)^{2}
+ \dfrac{\lambda_{2}}{2} \left(\Phi_{2}^{\dagger} \Phi_{2}\right)^{2}\notag\\
&+\enspace \lambda_{3} \Phi_{1}^{\dagger} \Phi_{1} \Phi_{2}^{\dagger} \Phi_{2}
+ \lambda_{4} \Phi_{1}^{\dagger} \Phi_{2} \Phi_{2}^{\dagger} \Phi_{1}
+ \dfrac{\lambda_{5}}{2} \left[\left(\Phi_{1}^{\dagger} \Phi_{2}\right)^{2} + \text{h.c.}\right]\label{eq:scalpot}\\
&+\enspace \dfrac{1}{2} m_{s}^{2}\Phi_{S}^{2} + \dfrac{\lambda_{6}}{8} \Phi_{S}^{4} + \dfrac{\lambda_{7}}{2} \Phi_{1}^{\dagger} \Phi_{1} \Phi_{S}^{2} + \dfrac{\lambda_{8}}{2} \Phi_{2}^{\dagger} \Phi_{2} \Phi_{S}^{2}\,,\notag
\end{align}
where $\Phi_i$ ($i=1,2$) represents scalar doublets, while the singelt field is denoted by $\Phi_S$. The potential has real parameters (mass terms and quartic couplings $\lambda_i$), and exact discrete symmetries. In particular, a $\mathbb{Z}^{(1)}_{2}$ symmetry under which all fields are even except for $\Phi_2$, and a $\mathbb{Z}^{(2)}_{2}$ symmetry under which all fields are even except for $\Phi_S$.

The VEVs of the fields are giving by: 
\begin{align}
\left<\Phi_1\right>=\begin{pmatrix}
0 \\ \frac{v_1}{\sqrt{2}}
\end{pmatrix},\qquad \left<\Phi_2\right>=\begin{pmatrix}
0 \\ \frac{v_2}{\sqrt{2}}
\end{pmatrix},\qquad\left<\Phi_S\right>=v_s\,,
\label{eq:n2hdm}
\end{align}
and the fields can be parameterized as follows:
\begin{align}
\Phi_1 = \begin{pmatrix} \phi_{1}^{+} \\
\dfrac{1}{\sqrt{2}}\left(v_{1} + \rho_{1} + i\,\eta_{1}\right)
\end{pmatrix},\quad
\Phi_2 = \begin{pmatrix} \phi_{2}^{+} \\
\dfrac{1}{\sqrt{2}}\left(v_{2} + \rho_{2} + i\,\eta_{2}\right)
\end{pmatrix}, \quad
\Phi_S = v_{s} + \rho_{s}\,.
\label{eq:para}
\end{align}
where $\phi_{1,2}^+$ are charged complex gauge eigenstates,  $\rho_{1,2,s}$ are neutral CP-even gauge eigenstates, and $\eta_{1,2}$ are neutral CP-odd gauge eigenstates.

The minimization conditions of the potential lead to,
\begin{align}
m_{11}^{2} &= - \dfrac{1}{2}\,  \left(v_{1}^{2}\lambda_{1} + v_2^2 \left(\lambda_3+\lambda_4+\lambda_5\right) + v_s^2 \lambda_7\right) \label{eq:tadpole1} \\
m_{22}^{2} &= - \dfrac{1}{2}\,  \left(v_1^2 \left(\lambda_3+\lambda_4+\lambda_5\right) + v_{2}^{2}\lambda_{2}  + v_s^2 \lambda_8\right) \label{eq:tadpole2} \\
m_s^2\,\, &= - \dfrac{1}{2}\, \left(v_1^2\lambda_7 + v_2^2\lambda_8 + v_s^2\lambda_6\right) 
\end{align}

An interesting variant of the model is the DDP-N2HDM, which is defined as the case where only $\Phi_1$ and $\Phi_S$ acquire VEVs ($v=246$ GeV and $v_s$), while $\Phi_2$ remains inert (setting $v_2 = 0$ in the previous equations for the general N2HDM). This leads to a DM candidate coming from the inert doublet. The stability of the DM particle requires the preservation of the $\mathbb{Z}^{(1)}_{2}$ symmetry, while the other symmetry $\mathbb{Z}^{(2)}_{2}$ is broken by $\Phi_S$ when it obtains a non-zero VEV. Moreover, the mass eigenstates of the CP-even neutral Higgs bosons ($m_{H_1}$ and $m_{H_2}$) are derived through rotating the gauge eigenstates by a $3\times3$ matrix $R$ that depends on the rotation angle $\alpha$ and whose components take the values: $R_{11} = R_{23}= \cos{\alpha}$, $R_{13} = - R_{21} = \sin{\alpha}$, $R_{32} = 1$, while the other components are zero. The CP-odd and charged gauge eigenstates of the two doublets do not mix. Therefore, the mass eigenstates coming from the inert doublet are part of the dark sector, and hence denoted by $m_{H^{\pm}_D}$, $m_{A_D}$, and ${m_{H_D}}$ (the DM candidate). 

The Yukawa Lagrangian is taken to be of type I in order to allow for comparison of different phases of the N2HDM. For the DDP of the N2HDM it is given as follows: 
\begin{align}
\mathcal{L}_{\text{Yukawa}} = -\bar{Q}^T_L Y_{U}\widetilde{\Phi}_1 U_R -\bar{Q}^T_L Y_{D}\Phi_1 D_R -\bar{L}^T_L Y_{L}\Phi_1 E_R + \text{h.c.}\,,
\end{align}
where $Y$ are Yukawa coupling matricese, $Q_L$ and $L_L$ are doublets containing the left-handed fermions, while $U_R, D_R$ and $E_R$ encompass right-handed fermions.

The quartic couplings \(\lambda_i\) can be expressed in terms of scalar masses, mixing angles, and VEVs \cite{Engeln:2020fld}:

\begin{align}
    \lambda_1 &= \frac{1}{v^2} \left( m_{H_1}^2 R_{11}^2 + m_{H_2}^2 R_{21}^2 \right), \\
    \lambda_3 &= \frac{1}{v^2} \left( 2m_{H^\pm_D}^2 - 2m_{22}^2 - v_s^2 \lambda_8 \right), \\
    \lambda_4 &= \frac{1}{v^2} \left( m_{A_D}^2 + m_{H_D}^2 - 2m_{H^\pm_D}^2 \right), \\
    \lambda_5 &= \frac{1}{v^2} \left( m_{H_D}^2 - m_{A_D}^2 \right), \\
    \lambda_6 &= \frac{1}{v_s^2} \left( m_{H_1}^2 R_{13}^2 + m_{H_2}^2 R_{23}^2 \right), \\
    \lambda_7 &= \frac{1}{vv_s} \left( m_{H_1}^2 R_{11} R_{13} + m_{H_2}^2 R_{21} R_{23} \right),
    \label{lambdas}
\end{align}
where \(R_{ij}\) are elements of the mixing matrix.

Finally, using the minimization conditions and the relations in Eq.\ref{lambdas}, the model can be specified by the following input parameters: $m_{H_1}, m_{H_2}, m_{H_D}, m_{A_D}, m_{H^{\pm}_D}$, $m^2_{22}$,  $v_s$, $\lambda_2$, $\lambda_8$, $\alpha$.

\subsection{The Constraints} \label{sunsec:constraints}

\subsubsection*{Boundedness from Below }

A crucial condition for a stable vacuum, meaning that the global minimum is associated with the electroweak symmetry breaking vacuum, is that the scalar potential must be bounded from below (\textit{BfB}). The potential must remain positive in the limit where the fields reach infinity. The potential is BfB if one of the following conditions is satisfied \cite{Klimenko:1984qx, Muhlleitner:2016mzt }:
\begin{itemize}
    \item The first set of conditions is:
\end{itemize}
\begin{align*}
\lambda_1 > 0, \ \lambda_2 > 0, \ \lambda_6 > 0 &, \\
\sqrt{\lambda_1 \lambda_6} + \lambda_7 > 0 &, \\
\sqrt{\lambda_2 \lambda_6} + \lambda_8 > 0 &, \\
\sqrt{\lambda_1 \lambda_2} + \lambda_3 + D > 0 &, \\
\lambda_7 + \sqrt{\frac{\lambda_1}{\lambda_2}} \lambda_8 \geq 0 &,
\end{align*}
where \( D = \lambda_4 - \lambda_5 \) if \(\lambda_4 > \lambda_5 \) and zero otherwise. 

\begin{itemize}
    \item The second set of conditions is:
\end{itemize}
\begin{align*}
\lambda_1 > 0, \ \lambda_2  > 0, \ \lambda_6 > 0 &, \\
\sqrt{\lambda_1 \lambda_6} > -\lambda_7 \geq \sqrt{\frac{\lambda_1}{\lambda_2}} \lambda_8 &, \\
\sqrt{\lambda_2 \lambda_6} \geq \lambda_8 > -\sqrt{\lambda_2 \lambda_6} &, \\
\sqrt{\left(\lambda_7^2 - \lambda_1 \lambda_6\right)\left(\lambda_8^2 - \lambda_2 \lambda_6\right)} > \lambda_7 \lambda_8 - (D + \lambda_3) \lambda_6 &.
\end{align*}

\subsubsection*{Perturbative Unitarity}

To ensure tree-level Perturbative Unitarity (\textit{PU}) of the model, one imposes an upper limit of $8\pi$ \cite{Horejsi:2005da} on the largest eigenvalue of the tree level $2 \to 2$ scattering matrix $\mathcal{M}_{2\to2}$. For the N2HDM, accounting for the contributions from the additional singlet field, the full expression was computed in \cite{Muhlleitner:2016mzt}, and the following conditions must apply,
\begin{equation}
\begin{aligned}
|\lambda_3 - \lambda_4| & \\
|\lambda_3 + 2 \lambda_4 \pm 3 \lambda_5| & \\
\left| \frac{1}{2} \left( \lambda_1 + \lambda_2 + \sqrt{(\lambda_1 -
     \lambda_2)^2 + 4 \lambda_4^2}\right) \right| &\\
\left| \frac{1}{2} \left( \lambda_1 + \lambda_2 + \sqrt{(\lambda_1 -
     \lambda_2)^2 + 4 \lambda_5^2}\right) \right| & \\
|\lambda_7|\,,\;|\lambda_8| & \\
\frac{1}{2}|a_{1,2,3}| & 
\end{aligned}
\left.\phantom{\frac{1}{2}}\right\} < 8\pi.
\label{eq:ev}
\end{equation}
where $a_{1,2,3}$ are given by \cite{Muhlleitner:2016mzt},
\begin{align}
a_1 &= 4\left(-27 \lambda_1 \lambda_2 \lambda_6 + 12 \lambda_3^2 \lambda_6 + 12
\lambda_3 \lambda_4 \lambda_6 + 3 \lambda_4^2 \lambda_6 + 6 \lambda_2
\lambda_7^2 - 8 \lambda_3 \lambda_7 \lambda_8 - 4 \lambda_4 \lambda_7
\lambda_8 + 6 \lambda_1 \lambda_8^2 \right), \\
a_2 &= 36 \lambda_1 \lambda_2 -
16\lambda_3^2 - 16\lambda_3 \lambda_4 - 4 \lambda_4^2 + 18 \lambda_1
\lambda_6 + 18 \lambda_2 \lambda_6 - 4\lambda_7^2 - 4\lambda_8^2, \\
a_3 &= -6 (\lambda_1 + \lambda_2) -3 \lambda_6.
\end{align}

\subsubsection*{Oblique  parameters}
Weak interaction observables can be indirectly affected by corrections to self-energies of gauge bosons arising from new physics. These indirect effects can be parameterized via the so-called oblique parameters: \( S \), \( T \), and \( U \) (referred to as \textit{STU})\cite{Peskin:1991sw}. It is possible to constrain these parameters via precision electroweak measurements.  
For multi-Higgs doublet and singlet models, the oblique parameters can be computed as shown in Refs. \cite{Grimus:2007if, Grimus:2008nb}. Their values, fitted in Ref. \cite{Haller:2018nnx}, are listed in Table \ref{tab:oblique_parameters}.
\begin{table}[h!]
    \centering
    \begin{tabular}{|c|c|c|c|}
        \hline
        Parameter & Value & Error & Correlation Coefficient \\ \hline
        \( S \) & 0.04 & \(\pm 0.11\) & \multirow{2}{*}{+0.92 (S and T)} \\ \cline{1-3}
        \( T \) & 0.09 & \(\pm 0.14\) & \\ \hline
        \( U \) & -0.02 & \(\pm 0.11\) & -0.68 (S and U), -0.87 (T and U) \\ \hline

    \end{tabular}
    \caption{Values of the oblique parameters \( S \), \( T \), and \( U \) with corresponding errors and correlation coefficients.}
    \label{tab:oblique_parameters}
\end{table}

\subsubsection*{Higgs Constraints}

The Higgs sector of the model is constrained by checking its compatibility with data from various collider experiments (for reviews, see Refs.~\cite{CMS:2022dwd, ATLAS:2022vkf}). The properties include Higgs couplings, branching ratios, production cross-sections, and signal rates. After computing the predictions of the model for these quantities, a comparison with experimental results is performed using the tools specified in Section 3. Namely, the branching ratios $BR(H_i \to X)$ of neutral and charged Higgs bosons $H_i$ into various SM particles $X$ are computed and checked against Higgs Data. Additionally, the production cross-sections of the Higgs bosons in various channels, such as gluon-gluon fusion ($gg \to H$), are calculated to check the resulting signal rates against the LHC observations. Finally, charged Higgs bosons are required to be heavier than 70 GeV \cite{Pierce:2007ut, Belyaev:2016lok}.

\section{Overview of multilabel Classification}
There are various types of classification problems. The most common is single-label classification, in which a set of features (e.g. the input parameters of a physics model) is associated with one target label, $L$. The label can be either binary (0 or 1) or multiclass (A, B, C, $\dots$). The aim of AI is to assign a class to each instance of features. This also applies to regression tasks. However, the goal of such tasks is to predict continuous or discrete numbers (range). In contrast, MLC is concerned with predicting more than one label given a set of features. Each label can be either binary (e.g. $L_1 \in \{0, 1\}$, $L_2 \in \{0, 1\}$, $\dots$) or multiclass (e.g. $L_1 \in \{a, b, c\}$, $L_2 \in \{a, b, c\}, \dots$). As for multi-output regression, the goal is to predict numerical values for each label.

It is possible to convert an MLC problem into a single-label multiclass problem. In this case, the classes of a single-label represent each possible combination of classes of multiple labels. For example, $n$ binary labels can be cast as a single multiclass label: $L \in \{(L_1=1 \ \text{AND} \ L_2=1), (L_1=1 \ \text{AND} \ L_2=0), \dots\}$, where the dots here represent the rest of possible $2^n$ combinations. However, it is important to note that, this practice could complicate the problem as the number of labels grows, since one would have to deal with each combination of classes separately. Moreover, the focus of this single-label multiclass classifier would be on the possible combinations only. It will not focus on each label individually, which could be problematic if one of the goals is to learn each individual label separately. With that in mind, MLC has the advantage of not only learning each label individually, but also the combinations of labels collectively, capturing any interrelations among them. We believe that MLC can be a very useful approach for studying parameter constraints in BSM extensions, especially if the aim is to gain detailed insights into the physics model.

\section{From Data to Deep Learning}
In this Section, we describe the methods used for data generation, handling class imbalance, optimizatin and fine-tuning, DL model building, evaluation metrics, and the ML baseline model. These methods are the base on which our \texttt{HEPMLC} tool is created. The tool is described in Appendix.

\subsection*{Data generation and class imbalance}

One of the most important (and often challenging) aspects of applying ML/DL is to generate good quality data, where there is no severe class imbalance. We use a hybrid method to sample the parameter space of the DDP-N2HDM. Specifically, a combination of Latin Hypercube Sampling (LHS) and random sampling was employed, and the final dataset comprises 774,262 points, ensuring good representation of the parameter space. In both sampling methods, multiple independent scans with different seeds were conducted to ensure that the samples did not cluster around specific regions. Furthermore, we restrict our analysis to the normal hierarchy variant where $m_{H_1} = 125.09$ GeV, while $m_{H_2} > m_{H_1}+3$ GeV. The scans span the ranges listed in Table~\ref{tab:rang}.

\begin{table}[ht]
    \centering
    \begin{tabular}{lll}
        \textbf{Parameter} & \textbf{Range} & \textbf{Description} \\
        \hline
        $m_{H_1}$ & 125.09 GeV & Mass of the first Higgs boson (SM-like) \\
        $m_{H_2}$ & 128 to 1500 GeV & Mass of the second Higgs boson \\
        $m_{H_D}$ & 1 to 1500 GeV & Mass of the DM candidate \\
        $m_{A_D}$ & 1 to 1500 GeV & Mass of the dark pseudoscalar \\
        $m_{H_D^{\pm}}$ & 1 to 1500 GeV & Mass of the charged dark Higgs boson \\
        $\alpha$ & -1.57 to 1.57 & Mixing angle \\
        $m_{22}^2$ & $10^{-3}$ to $5 \times 10^5$ GeV$^2$ & mass-squared parameter of $\Phi_2$ \\
        $\lambda_2$ & 0 to 20 & Quartic coupling constant \\
        $\lambda_8$ & -30 to 30 & Quartic coupling constant \\
        $v_s$ & 1 to 1500 GeV & Vacuum expectation value of the singlet field \\
    \end{tabular}
    \caption{Parameter ranges and descriptions used for Latin Hypercube Sampling and Random Sampling.}
    \label{tab:rang}
\end{table}

For each generated point, \texttt{ScannerS} \cite{Muhlleitner:2020wwk, Engeln:2018mbg} is utilized to compute the spectrum of the model, which includes physical masses, couplings, and several predicted quantities such as branching ratios, cross-sections, and quantities relevant to constraints, such as the maximum eigenvalue of the 2 by 2 S-matrix and the oblique parameters. A comprehensive check of theoretical and experimental constraints is carried out using internal routines and interfaced tools. In particular, the theoretical checks include \textit{BfB} and \textit{PU}, while the experimental ones cover \textit{STU} and \textit{Higgs} constraints. The \textit{STU} parameters are computed using the formulas mentioned in Section 2.2 and subsequently checked in \texttt{ScannerS} against their experimentally fitted values given in \cite{Haller:2018nnx}. On the other hand, the \textit{Higgs} constraints are checked using \texttt{HiggsBounds 5.10.2} \cite{Bechtle:2020pkv} and \texttt{HiggsSignals 2.6.2} \cite{Bechtle:2020uwn}.

The generated dataset contains the four aforementioned constraints as target labels, which are grouped into $16$ different combinations (i.e., valid/invalid \textit{BfB}, valid/invalid \textit{PU}, valid/invalid \textit{STU}, valid/invalid \textit{Higgs}). 

\begin{table}[h!]
    \centering
    \begin{tabular}{lccc}
        \toprule
        \textbf{Label} & \textbf{Invalid (0)} & \textbf{Valid (1)} & \textbf{Imbalance Ratio} \\
        \midrule
        BFB   & 308,111 & 466,151 & 1.513 \\
        PU   & 389,253 & 385,009 & 0.989 \\
        STU   & 341,024 & 433,238 & 1.270 \\
        Higgs & 328,299 & 445,963 & 1.358 \\
        \bottomrule
    \end{tabular}
    \caption{Class imbalance for individual labels in the dataset.}
    \label{tab:class_imbalance}
\end{table}

\begin{table}[h!]
    \centering
    \begin{tabular}{llll}
        \toprule
        \multicolumn{4}{c}{\small{BfB\_PU\_STU\_Higgs} (number of points in the dataset)} \\
        \midrule
        1\_1\_1\_1 (152974) & 0\_0\_0\_0 (91361) & 1\_0\_1\_1 (75904) & 1\_0\_0\_1 (46122) \\
        0\_1\_1\_1 (40367) & 1\_0\_1\_0 (39611) & 1\_0\_0\_0 (39329) & 1\_1\_0\_1 (39017) \\
        1\_1\_1\_0 (37868) & 0\_0\_1\_0 (35613) & 1\_1\_0\_0 (35326) & 0\_0\_0\_1 (34070) \\
        0\_1\_0\_1 (30266) & 0\_0\_1\_1 (27243) & 0\_1\_0\_0 (25533) & 0\_1\_1\_0 (23658) \\
        \bottomrule
    \end{tabular}
    \caption{Class combinations and their counts in the dataset.}
    \label{tab:class_combinations}
\end{table}

The final dataset was accepted after we ensured that each label was not severely imbalanced, as shown in Table~\ref{tab:class_imbalance}. The valid/invalid ratio ranges from about 1 (\textit{PU}) to 1.5 (\textit{BFB}), which is not severe. Furthermore, Table~\ref{tab:class_combinations} shows the class imbalance in the joint class distribution, and we ensured that each case was represented by sufficient points to avoid severe joint class imbalance. This is a crucial step for MLC tasks, since significant imbalance in individual labels would negatively affect model training overall, while imbalance in the joint class distributions would negatively affect the ability of the DL model to learn the interplay between labels.

We addressed severe imbalance at the initial stages of data generation by performing targeted scans using the physics tool, thereby populating underrepresented regions. By doing so, the DL model is exposed to a more comprehensive range of multilabel patterns, enabling it to learn the underlying joint distribution of constraints rather than only the most frequent patterns.

In the final dataset, a noticeable (not severe) class imbalance remains between the top \((1\_1\_1\_1)\) and bottom \((0\_1\_1\_0)\) joint classes, with a ratio of approximately 6:1. Also, since any combination containing at least one zero is overall invalid, we have a total of $152,974$ valid cases and $621,288$ invalid cases, corresponding to a 1:4 ratio. In practice, completely avoiding class imbalance is unrealistic for BSM models, as large regions of parameter space are expected to violate one or more constraints. We will address moderate class imbalance during training by selecting an appropriate loss function, which we will discuss later.

For completeness, we checked the vacuum structure using \texttt{EVADE} \cite{evade, Hollik:2018wrr, Ferreira:2019iqb} and DM relic density (upper limit) and direct detection using \texttt{MicrOMEGAs 5.3.41} \cite{Belanger:2020gnr}. However, we do not consider these as target labels in training our DL model since we observe that they bring about severe class imbalance in the dataset (hendering the quality of the data), with the majority class being valid (i.e. both are valid in the majority of the parameter space). Handling this is possible, but would exponentially increase the number of regions that needs to be populated from 16 (for 4 labels) to 64 (for six labels). This task is suitable for an analysis that utilizes high-performance computing and opens the door for exploring optimization techniques for fast sampling specific to MLC tasks, which we leave for a future work. Moreover, in the final dataset, valid regions with respect to the four target labels are also valid with respect to vacuum stability, and are almost entirely valid with respect to DM, expect for a very small region where $m_{DM} \leq 100$ GeV. Since our focus is on creating a multilabel classifier for the four target labels specified previously, we will not discuss vacuum stability and DM any further.

In terms of data preprocessing, it is generally crucial to analyze and prepare the dataset carefully before passing it to the DL model. This includes an initial analysis of the distribution of input parameters (features) listed in Table~\ref{tab:rang}, and the class imbalance of the target label(s). In our case, we consider the standardization of features by applying (if necessary) a Yeo-Johnson (YJ) power transformation \cite{yeo2000new}, which is suitable since some of the features take negative values (namely $\alpha$ and $\lambda_8$), followed by scaling (i.e. applying z-score normalization). The data is then split into 3 disjoint sets for training ($70\%$), validation ($15\%$), and testing ($15\%$). These tasks are carried out using  \texttt{Scikit-learn} \cite{pedregosa2011scikit}.

\subsection*{DL Model Architecture and Training}
Creating a deep neural network model for MLC involves a number of steps, such as selecting appropriate values of hyperparameters, constructing the network architecture, and configuring the optimization process. In this work, the creation of the DL model is designed to support both static and dynamic hyperparameter setup as will be further explained shortly below. The construction is done using \texttt{Tensorflow} \cite{abadi2016tensorflow}, and \texttt{Keras} \cite{keras}. A thorough experimentation phase is carried out by utilizing the \texttt{Optuna} framework \cite{akiba2019Optuna}, to obtain the best structure and values for the following:

\begin{itemize}
    \item \textbf{Number of Layers ($n_{\text{layers}}$):} The number of dense layers in the network, chosen between 2 and 3\footnote{We have experimented with larger numbers of hidden layers and found no significant improvements compared to using 3 hidden layers.}.
    \item \textbf{Number of Units per Layer ($n_{\text{units}, i}$):} The number of neurons in each layer, with values ranging from 128 to 1024.
    \item \textbf{Activation Function:} The activation function used in each layer, either ReLU \cite{nair2010rectified} or Leaky ReLU \cite{maas2013rectifier}.
    \item \textbf{Dropout Rate:} The dropout rate \cite{srivastava2014dropout} applied after each dense layer, ranging from 0.1 to 0.5.
    \item \textbf{Batch Normalization:} A boolean flag indicating whether batch normalization \cite{ioffe2015batch} is applied after each dense layer.
    \item \textbf{Optimizer:} The optimization algorithm, either Adam \cite{kingma2014adam} or Nadam \cite{dozat2016incorporating}.
    \item \textbf{Learning Rate:} The learning rate for the optimizer, selected on a logarithmic scale between $1 \times 10^{-5}$ and $1 \times 10^{-2}$.

\end{itemize}

It is important to note that Batch Normalization and  Dropout are introduced to prevent overfitting. 
The network is finalized with an output layer containing one neuron per label, using the sigmoid activation function \cite{dubey2022activation}.

\subsection*{Model Optimization and Evaluation}
The model is compiled using focal binary cross-entropy loss function \cite{lin2017focal}, 
\begin{equation}
    \mathcal{L}_{\text{focal}}(y,p) = - \Bigl[\,y \,(1 - p)^\gamma \,\log(p) \;+\; (1 - y)\,p^\gamma \,\log(1 - p)\Bigr]
\end{equation}
where $p$ is the predicted probability, $y \in \{0,1\}$ is the ground-truth label, and $\gamma$ is the focusing parameter (default value 2). This loss function pays special attention to hard examples that may be misclassified by the DL model, especially in the presence of class imbalance, by reducing the weight for cases that are well-classified.

 During the training and validation stage, a number of performance-measuring metrics, which are appropriate for MLC, are monitored to assess the DL model. These metrics include (for a review, see Ref. \cite{grandini2020metrics}):
    
\begin{itemize}
    \item \textbf{Subset Accuracy}: This is a distinct metric for MLC that considers an instance as correctly classified only if all labels are correctly predicted. For the entire dataset, it takes values between 0 (at least one label wrongly predicted for all instances) and 1 (all labels correctly predicted for all instances). We judge the success of our DL model based on this rather strict metric.
    
    \begin{equation}
    \text{Subset Accuracy} = \frac{1}{N} \sum_{i=1}^N \mathbf{1}\Bigl(\hat{Y}_i = Y_i\Bigr)
    \end{equation}
    \noindent
    where \(N\) is the number of samples, \(i\) indexes each sample (\(i=1,\ldots,N\)), \(\mathbf{1}(\cdot)\) is the indicator function (\(\mathbf{1}(\text{true})=1\), \(\mathbf{1}(\text{false})=0\)), \(\hat{Y}_i\) is the predicted (multilabel) set for the \(i\)-th sample, and \(Y_i\) is the ground-truth (multilabel) set for the \(i\)-th sample.

    \item \textbf{Hamming Loss}: This is a less strict metric representing the fraction of labels that are incorrectly predicted across all instances. This metric does not quantify whether all labels are incorrectly predicted for each instance. It takes values between 0 (all predictions correct) and 1 (all predictions incorrect).
    
    \begin{equation}
    \text{Hamming Loss} = \frac{1}{N \times L} \sum_{i=1}^{N} \sum_{j=1}^{L} \mathbf{1}\Bigl(y_{ij} \neq \hat{y}_{ij}\Bigr)
    \end{equation}
    \noindent
    where \(N\) is the number of samples, \(L\) is the number of labels per sample, \(i\) indexes each sample, \(j\) indexes each label (\(j=1,\ldots,L\)), \(\mathbf{1}(\cdot)\) is the indicator function, \(y_{ij}\) is the ground-truth label, and \(\hat{y}_{ij}\) is the predicted label for label \(j\) of sample \(i\).

    \item \textbf{Matthews Correlation Coefficient (MCC)}: This metric takes into consideration true positives (TP), true negatives (TN), false positives (FP), and false negatives (FN). It combines them into a single formula and outputs 1 for perfect predictions, 0 for random classification, and -1 for completely wrong predictions.
    
    \begin{equation}
    \text{MCC} = \frac{TP \times TN - FP \times FN}{\sqrt{(TP + FP)(TP + FN)(TN + FP)(TN + FN)}}
    \end{equation}

    \item \textbf{Macro F1-Score}: This is the average of the F1-Scores calculated for each label independently. The F1-Score is the harmonic mean of Precision (\(TP/(TP + FP)\)) and Recall (\(TP/(TP+FN)\)).
    
    \begin{equation}
    \text{F1}_j = \frac{2 \times \text{Precision}_j \times \text{Recall}_j}{\text{Precision}_j + \text{Recall}_j},
    \quad
    \text{Macro-F1} = \frac{1}{L} \sum_{j=1}^{L} \text{F1}_j
    \end{equation}
    where \(L\) is the number of labels, \(j\) indexes the labels.

    \item \textbf{Accuracy}: This is used only when assessing each label individually, and represents the number of correct predictions over the total number of predictions.
    
    \begin{equation}
    \text{Accuracy} = \frac{TP + TN}{TP + TN + FP + FN}
    \end{equation}
    \noindent
    where \(TP\), \(TN\), \(FP\), and \(FN\) are defined as above.

    \item \textbf{Confusion Matrix}: This matrix illustrates the number of TP, TN, FP, and FN for each label, providing detailed insight into where the DL model performs well and where it performs poorly.

\end{itemize}

In addition to these standard MLC metrics, we can further evaluate the capacity of the model to learn the unique class combinations (Table~\ref{tab:class_combinations}) by conducting a powerset evaluation. Namely, after the DL model is trained and labels are predicted, one treats each combination of labels as a separate class and computes precision, recall, and F1-scores. While subset accuracy is already a strict metric, it can be influenced by a few dominant combinations. By examining each combination individually, the powerset evaluation tells us if the model truly performs well across the entire range of label combinations, providing deeper insight into whether it has captured patterns between the combined labels.

In the process of optimizing our DL model, we introduce a custom score for \texttt{Optuna} to maximize during its trials. The score combines validation subset accuracy, validation loss, and validation MCC, and we define it as follows:

\[
\text{Custom Score} =
\left(\alpha \times \text{Val Subset Accuracy}\right) +
\left(\beta \times \frac{1 / (\text{Val Loss} + \epsilon)}{1 + 1 / (\text{Val Loss} + \epsilon)}\right) +
\left(\delta \times \frac{\text{Val MCC} + 1}{2}\right),
\]
where $\alpha$, $\beta$, and $\delta$ represent scaling coefficients for metric terms which are all themselves designed to range from 0 and 1. $\epsilon$ is a small constant to prevent division by zero. The reason for introducing this score is that focusing on validation loss alone during the \texttt{Optuna} trials can lead to a model that underperforms with respect to the other validation metrics that we consider crucial. Once the experimentation phase is completed and the best parameters are found, we retrain the model using those parameters and apply early stopping based on minimizing the validation loss, which is common practice.

Finally, since to our knowledge there are no previous works in this context against which we can compare the performance of our DL model, we train a binary RF classifier \cite{pedregosa2011scikit} on the dataset as a baseline. The hyperparameters of the RF classifier are set to the default values given by \texttt{Scikit-learn}, which are: 100 decision trees, no restriction on the maximum depth, a minimum sample split of 2, a minimum sample leaf of 1, the Gini impurity criterion for the evaluation of splits, and the square root of the number of features as the number of features considered for each split. In addition, bootstrap sampling (bagging) was enabled to reduce overfitting\footnote{We did not observe any significant improvements from manually tuning the hyperparameters of the RF classifier.}.

\section{Results and Discussion} \label{sc:res}
In this Section, we present the results for our DL model and the physics analysis. For the DL model, we show the training history and the evaluation on a completely unseen test set comprising 116,014 points (instances). For the physics analysis, we use the correctly labeled instances from the test set and show, in the appropriate input $x-y$ plane, how the constraints affect the parameter space, both individually and collectively.

\subsection{Neural network architecture and training history}
Our DL model was trained on the following input parameters (features) and constraints (binary labels):
\begin{itemize}
    \item \textbf{Features:} $m_{H_2}$, $m_{H_D}$, $m_{A_D}$, $m_{H^{\pm}_D}$, $\alpha$, $\lambda_2$, $\lambda_8$, $v_s$, $m_{22}^2$
    \item \textbf{Labels:} BfB, PU, STU, Higgs
\end{itemize}

\begin{table}[]
\centering
\begin{tabular}{|l|l|}
\hline
\textbf{Parameter} & \textbf{Value} \\
\hline
Apply Yeo-Johnson Transformation & Yes \\
Apply Standard Scaler & Yes \\
Batch Size & 781 \\
Number of Layers & 3 \\
Units in Layer 1 & 875 \\
Units in Layer 2 & 938 \\
Units in Layer 3 & 402 \\
Activation Function & ReLU \\
Dropout Rate & 0.117 \\
Apply Batch Normalization & Yes \\
Optimizer & Adam \\
Learning Rate & 0.0003 \\

Custom Score weights\tablefootnote{These were experimented with manually, so they are not suggested by \texttt{Optuna}.} & $\alpha=\beta=0.4, \ \delta = 0.2$ \\
\hline
\end{tabular}
\caption{Optimized Parameters for the Deep Learning Model}
\label{tab:optimized_params}
\end{table} 

The architecture of the DL model and its hyperparameters were selected after experimenting with 100 \texttt{Optuna} trials, which are shown in Table \ref{tab:optimized_params}. Namely, we have an input layer that receives preprocessed features, where both YJ transformation and Standard scaling are applied. Next, we have 3 deep layers with 875 neurons for the first, 938 for the second, and 402 for the third. Each layer is followed by ReLU activation, then Batch Normalization and Dropout (0.117). We finally have an output layer comprising 4 neurons (one for each label), and a Sigmoid activation function. The training is optimized via the Adam optimization algorithm with a learning rate of 0.0003.

\begin{figure}[]
    \centering
    \begin{subfigure}[b]{0.45\textwidth}
        \centering
        \includegraphics[width=\textwidth]{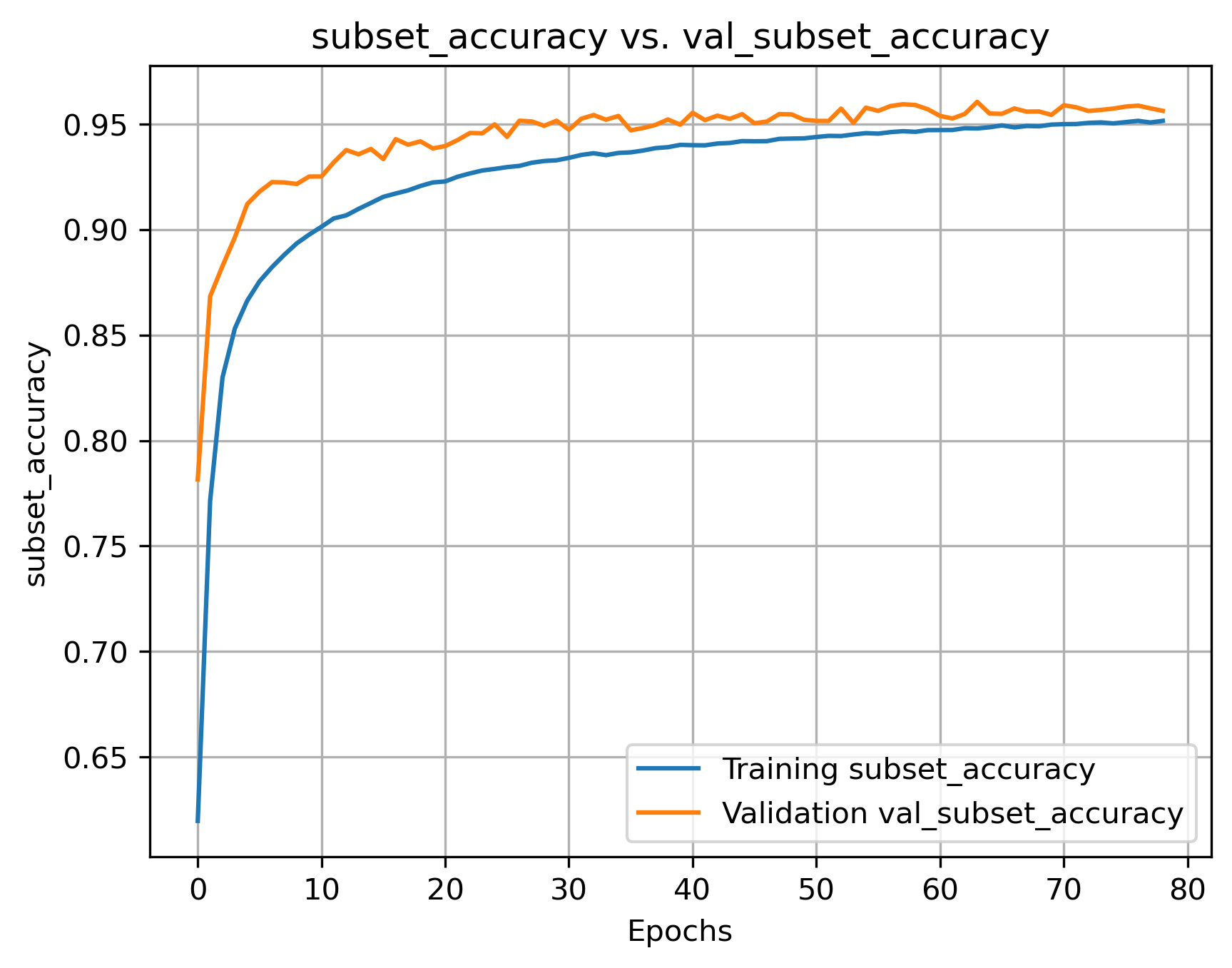}
        \caption{Subset Accuracy}
    \end{subfigure}
    \hfill
    \begin{subfigure}[b]{0.45\textwidth}
        \centering
        \includegraphics[width=\textwidth]{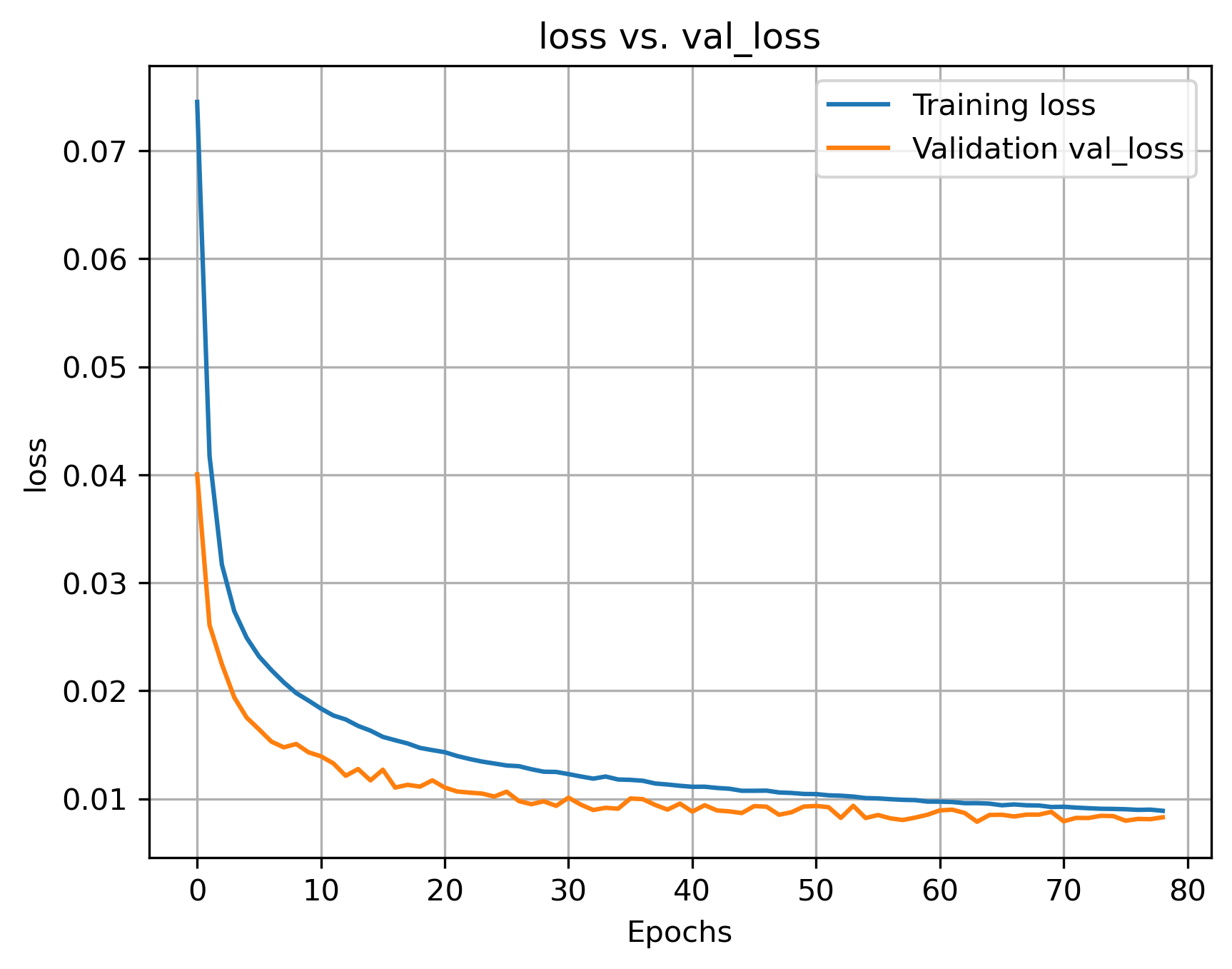}
        \caption{Loss vs. Validation Loss}
    \end{subfigure}
    
    \vspace{0.5cm}

    \begin{subfigure}[b]{0.45\textwidth}
        \centering
        \includegraphics[width=\textwidth]{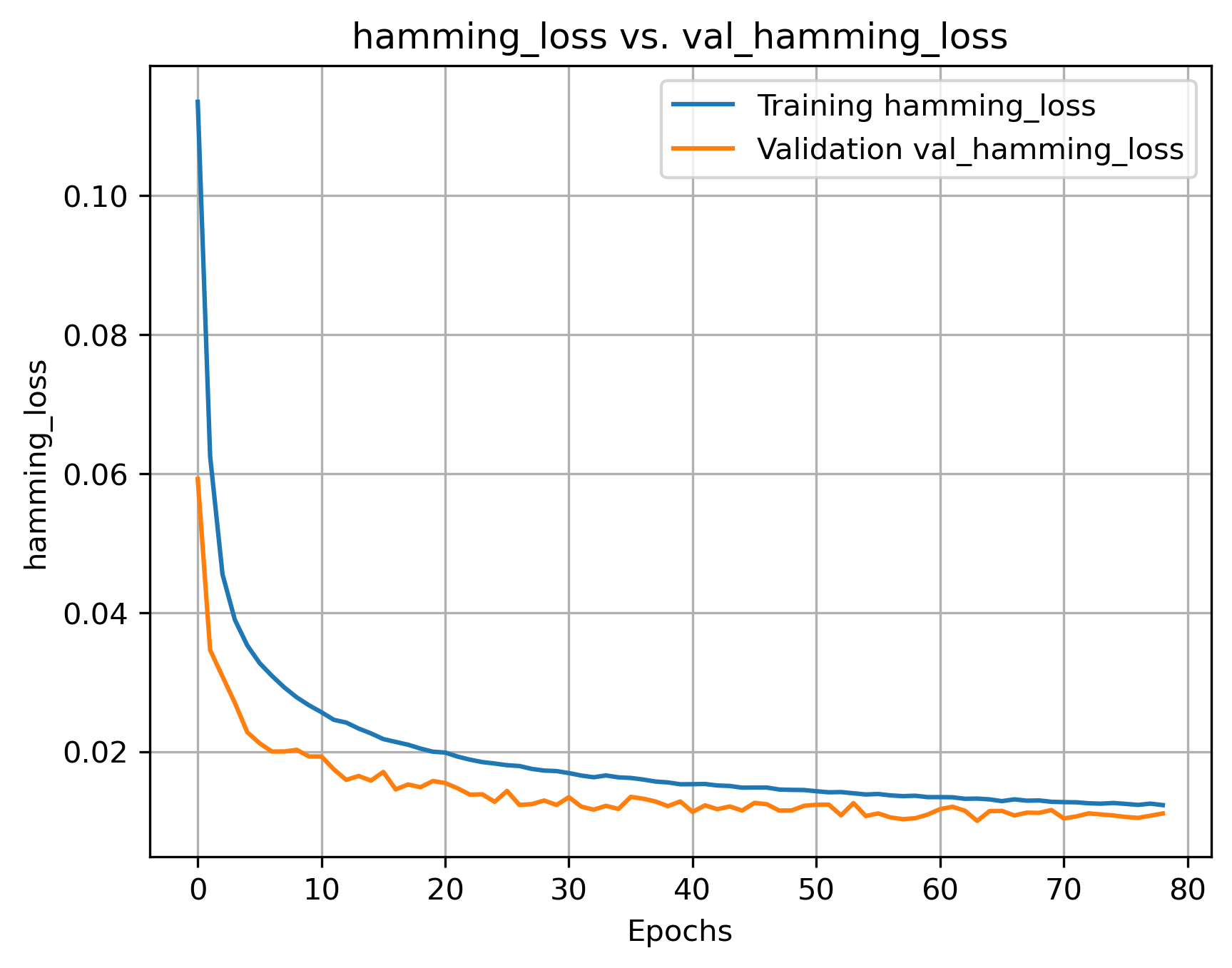}
        \caption{Hamming Loss}
    \end{subfigure}
    \hfill
    \begin{subfigure}[b]{0.45\textwidth}
        \centering
        \includegraphics[width=\textwidth]{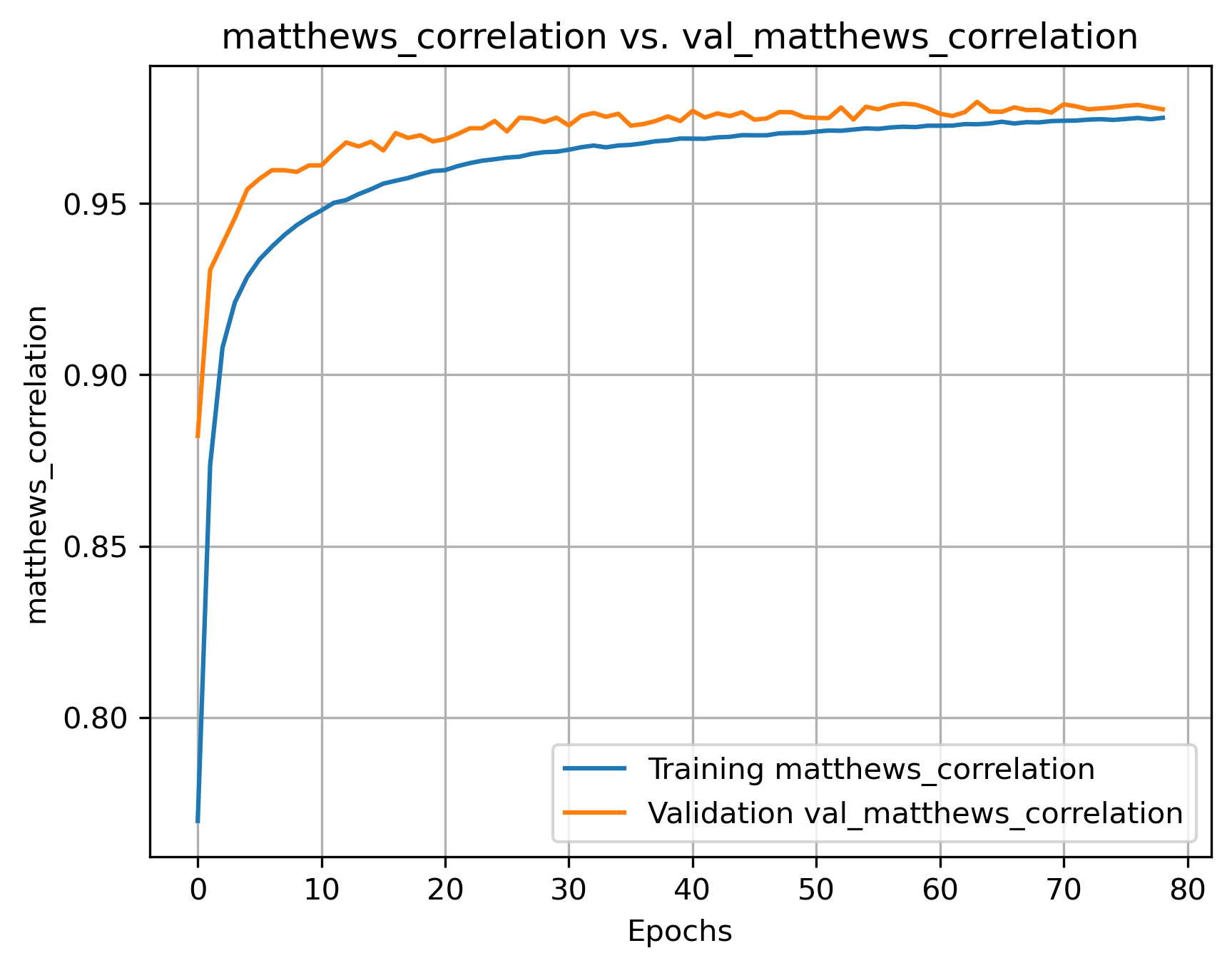}
        \caption{Matthews Correlation}
    \end{subfigure}
    
    \vspace{0.5cm}

    \begin{subfigure}[b]{0.45\textwidth}
        \centering
        \includegraphics[width=\textwidth]{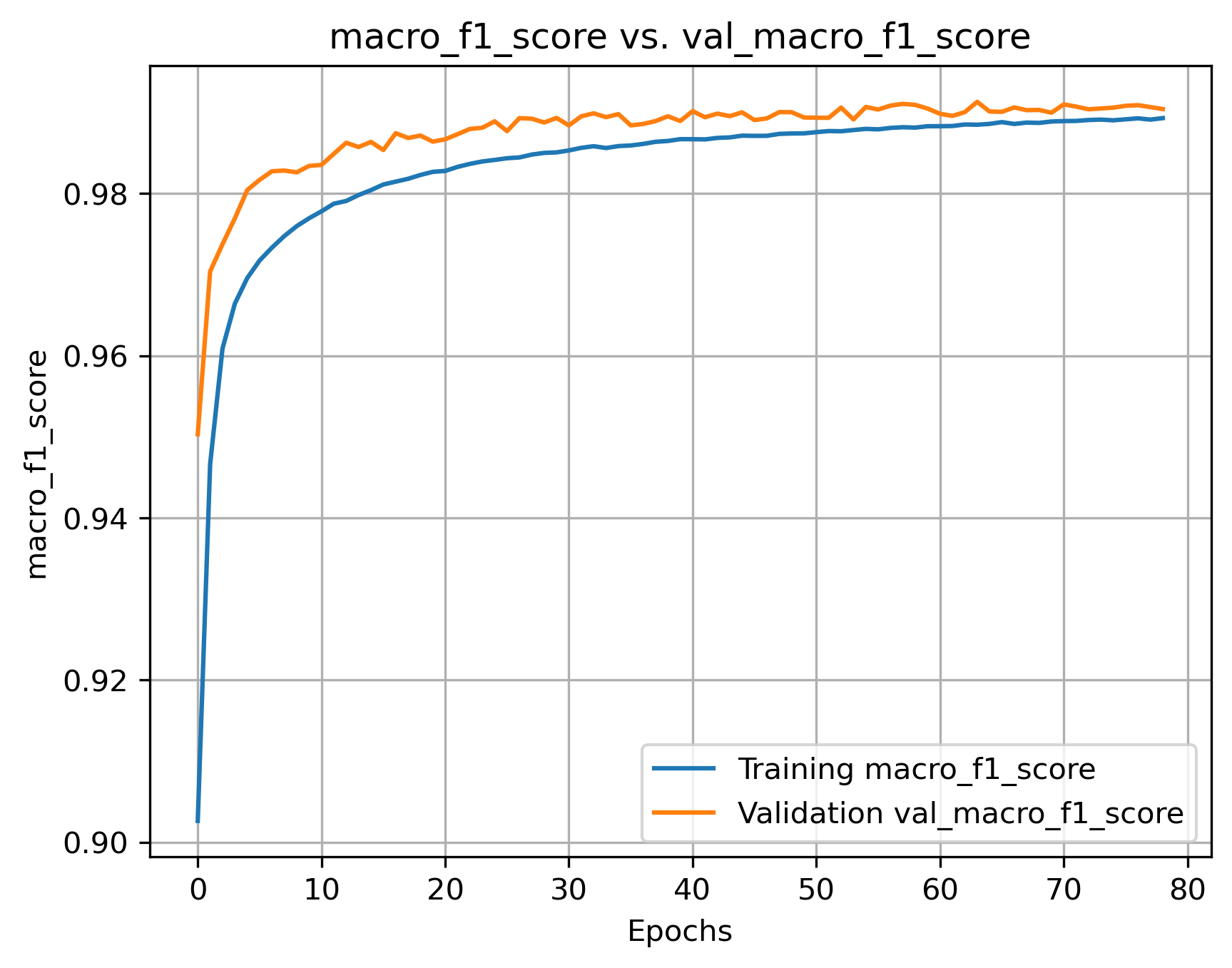}
        \caption{Macro F1 Score}
    \end{subfigure}
    \caption{Training history metrics for the model, including subset accuracy, loss, Hamming loss, Matthews correlation, and macro F1 score for both training and validation datasets.}
    \label{train-val-history}
\end{figure}

The training and validation histories are illustrated in Fig.\ref{train-val-history}, which presents the subset accuracy, loss, Hamming loss, MCC, and macro F1 score as functions of epochs. These plots provide crucial information regarding the performance of the DL model during the training stage, and can reveal if it suffers from overfitting.

Starting with Fig.\ref{train-val-history}-a), we can see the evolution of the subset accuracy, which we consider the ultimate measure of our DL model given that it focuses on all labels at once. The figure shows a rapid improvement in the ability to predict all labels correctly, with accuracy stabilizing around 0.95 after approximately 60 epochs. The gap between the training and validation shrinks as the training advances, which indicates to us that the model is
not overfitting, and is effectively learning global features the data. Moreover, the loss plot (Fig.\ref{train-val-history}-b) reveals a significant decrease in both training and validation loss during the early epochs, both plateauing at around 0.01 when epoch 80 is reached. This decrease in loss means that the DL model is able to minimize the difference between predicted and actual values, and the smallness of the gap at the end reassures us that the model is not overfitting. The Hamming loss, shown in Fig.\ref{train-val-history}-c, for both training and validation also decreases sharply in the early epochs and stabilizes at around 0.015. This reduction further confirms the ability of the DL model to make fewer incorrect predictions as training progresses. Next, the MCC plot (Fig.\ref{train-val-history}-d) shows an increasing improvement during the initial epochs, with values stabilizing around 0.96 after approximately 60 epochs. This suggests that the DL model learns to make balanced predictions (decreasing cases of FP and FN), with consistent performance as seen from the curves of both the training and validation datasets. Fig.\ref{train-val-history}-e, illustrates the macro-averaged F1 score and demonstrates a continuing improvement that stabilizes around 0.98. This high F1 score highlights that the DL model can balance precision and recall. The consistency between the training and validation in all of the metrics shown indicates that the DL model is robust and does not suffer from alarming overfitting.

\subsection{Performance evaluation}
To confirm whether the trained DL model can generalize to completely unseen data, we evaluate its performance on the test dataset (116,140 points). In particular, since the main task is to optimize the DL model for MLC, we consider the subset accuracy (on the test dataset) to be the major indicator of its performance. Its value is found to be 0.96, which means that it correctly predicted the values of all labels simultaneously of $96\%$ of the test dataset (111,504 points). To gain insight into how significant this result is, we compare the subset accuracy we obtained with that of an RF classifier, which we trained on the same dataset and evaluated on the same unseen test dataset. We find that the subset accuracy of the RF classifier in the test set is 0.78. Evidently, the DL multilabel classifier significantly outperforms the traditional RF classifier. 

\begin{figure}[H]
    \centering
    \begin{subfigure}[b]{0.45\textwidth}
        \centering
        \includegraphics[width=\textwidth]{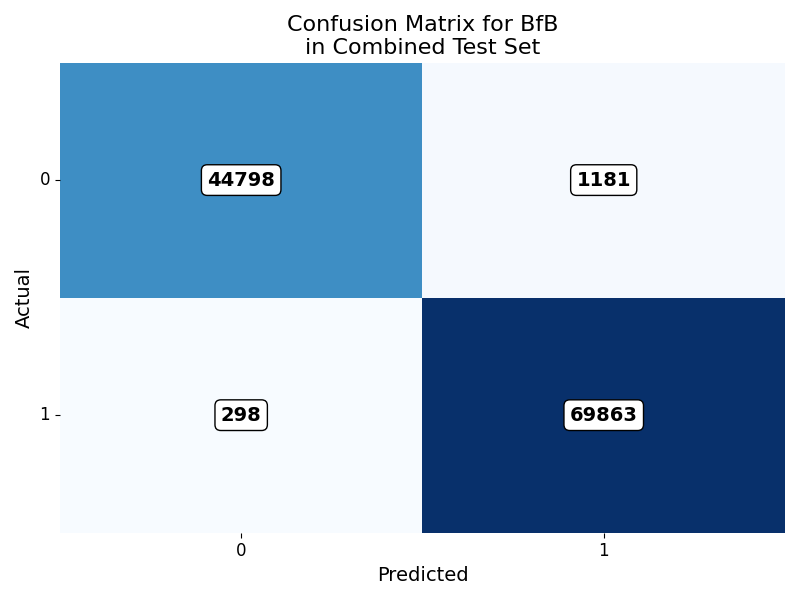}
        \caption{Confusion Matrix for BfB}
    \end{subfigure}
    \hfill
    \begin{subfigure}[b]{0.45\textwidth}
        \centering
        \includegraphics[width=\textwidth]{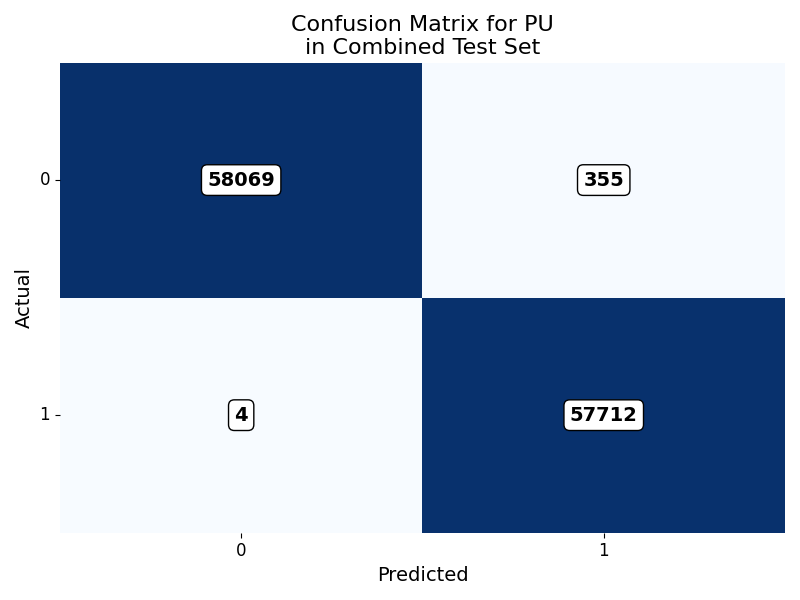}
        \caption{Confusion Matrix for PU}
    \end{subfigure}
    
    \vspace{0.5cm}

    \begin{subfigure}[b]{0.45\textwidth}
        \centering
        \includegraphics[width=\textwidth]{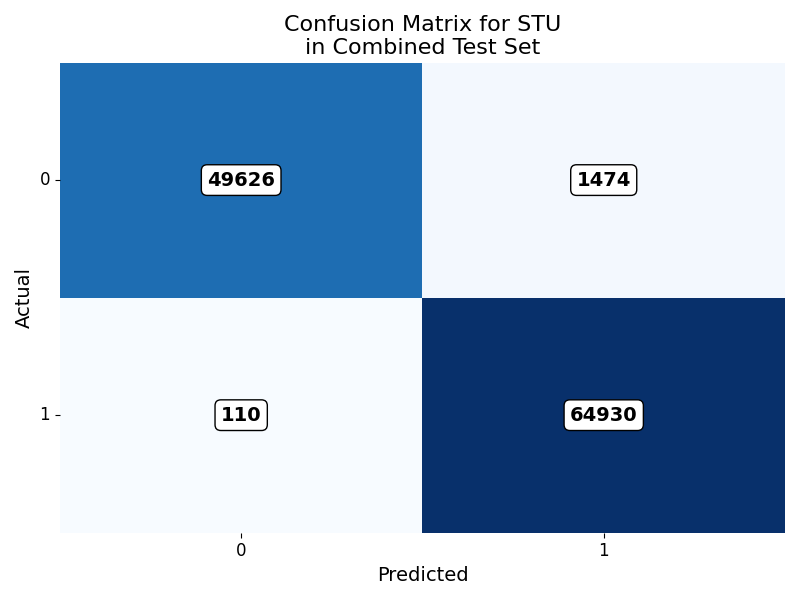}
        \caption{Confusion Matrix for STU}
    \end{subfigure}
    \hfill
    \begin{subfigure}[b]{0.45\textwidth}
        \centering
        \includegraphics[width=\textwidth]{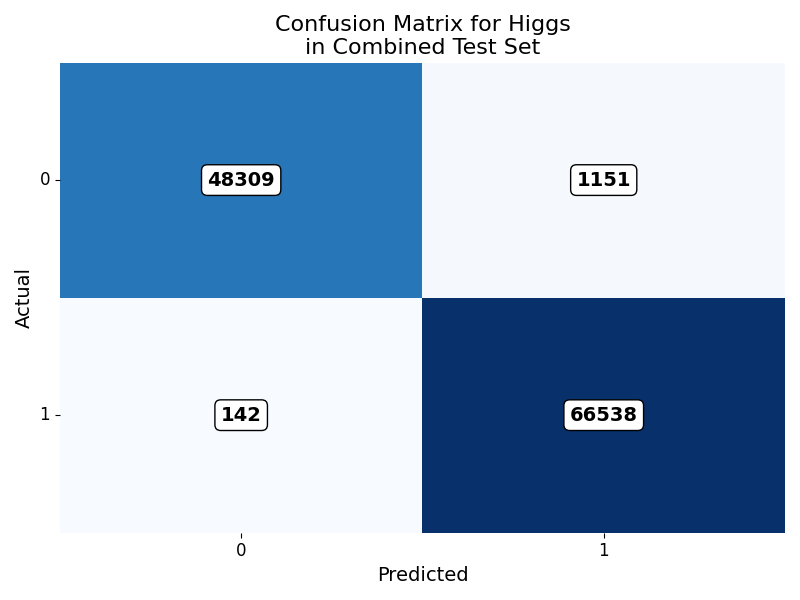}
        \caption{Confusion Matrix for Higgs}
    \end{subfigure}
    \caption{Confusion matrices for the BfB, PU, STU, and Higgs labels, showing the model's performance in predicting valid and invalid cases for each label separately.}
    \label{conf-matrices}
\end{figure}

Furthermore, it is vital to assess the performance of the DL model on individual labels, which will evaluate it as a single-label classifier, along with being used for MLC. This can be done by computing the confusion matrix for each label. The results are presented in Fig.~\ref{conf-matrices}. Each confusion matrix illustrates the model's ability to distinguish a valid case (class 1) from an invalid case (class 0). For \textit{BfB}, the confusion matrix (Fig.~\ref{conf-matrices}-a) shows that the DL model was able to correctly identify 69,863 valid cases and 44,798 invalid cases, with 1,181 FP and 298 FN predictions. As for \textit{PU}, Fig.~\ref{conf-matrices}-b illustrates only 4 FN and 355 FP misclassifications, while 58,069 cases were correctly classified as invalid and 57,712 correctly classified as valid. Next, for the \textit{STU} label, the DL model correctly predicted 64,930 valid cases and 49,626 invalid cases, as shown in Fig.~\ref{conf-matrices}-c. The figure also shows 1,474 FP and 110 FN cases. Finally, the confusion matrix for the \textit{Higgs} label (Fig.~\ref{conf-matrices}-d) shows that the DL model correctly identified 66,538 valid cases and 48,309 invalid cases, with 1,151 FP and 142 FN cases. These results highlight a very strong performance in the single-label classification task.

The performance of the DL model on individual labels can also be measured by computing accuracy (for individual labels, not to be confused with subset accuracy, which is relevant for MLC), precision, recall, and F1-score from the confusion matrices in Fig.~\ref{conf-matrices}. Starting with the \textit{BfB} label, the accuracy is 0.9872, with a precision of 0.9834, a recall of 0.9957, and an F1-score of 0.9895. As for the \textit{PU} label, the accuracy is 0.997, with a precision of 0.9939, a recall of 0.9999, and an F1-score of 0.9969. Next, the \textit{STU} label has an accuracy of 0.9864, with a precision of 0.9778, a recall of 0.9983, and an F1-score of 0.9879. Finally, the \textit{Higgs} label shows an accuracy of 0.989, with a precision of 0.983, a recall of 0.9979, and an F1-score of 0.9904. Overall, these metrics demonstrate strong performance across all labels, further indicating that our DL model can be effectively utilized for predicting each label individually (as a single-label classifier) as well as all labels at once (as a multilabel classifier).

\begin{figure}[ht]
    \centering
    \includegraphics[width=0.8\textwidth]{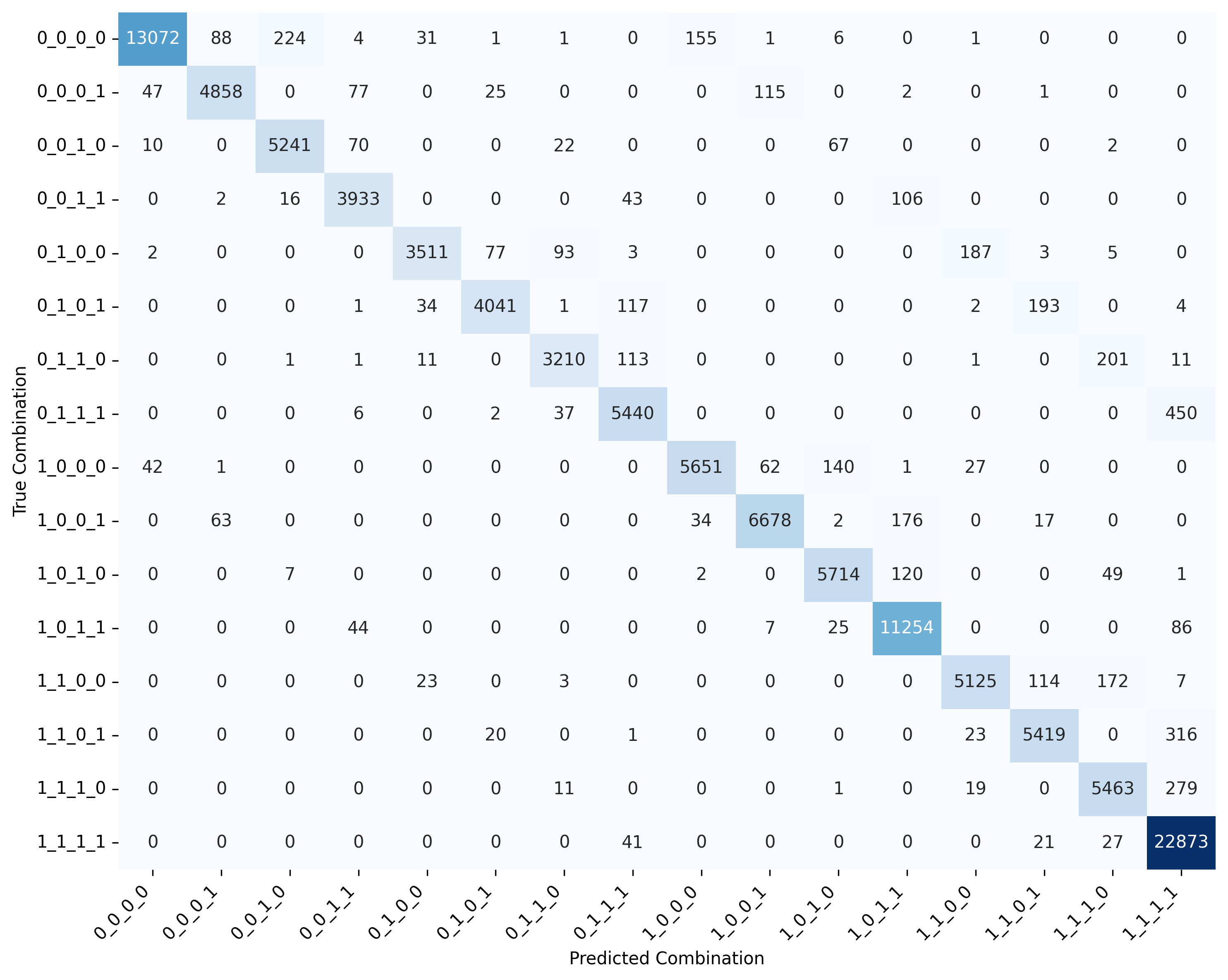}
    \caption{Powerset-based confusion matrix for the multilabel classifier. Each cell $(i,j)$ shows how often class $i$ (true) is predicted as class $j$.}
    \label{fig:powerset_conf_matrix}
\end{figure}

Finally, we further validate that our DL model captures the underlying joint distribution of labels (Table \ref{tab:class_combinations}). We treated each of the 16 possible label combinations as a single class and conducted a powerset  evaluation. Table \ref{tab:powerset_clf} presents the precision, recall, F1-score, and support for each of these combined classes, while the accuracy is the subset accuracy found before. Notably, all combinations achieve consistently high performance metrics ranging from F1-score of $\sim 93\%$ to $\sim 98\%$. Moreover, we can see from the confusion matrix in Figure \ref{fig:powerset_conf_matrix} that misclassification is rare. However, when the joint class $(1\_1\_1\_1)$ is misclassified, it is mainly misclassified as a joint class with at least three valid labels. The same observation is true for the joint class $(0\_0\_0\_0)$ and the other classes in which the misclassification arises due to one constraint. There are no caes in which a completely invalid result was identified as completely valid. These results strongly suggest that the model effectively learns and reproduces the joint distribution of the constraints, rather than simply focusing on the most frequent combinations, and provides reliable predictions.

\begin{table}[h!]
\centering
\begin{tabular}{c c c c c}
\toprule
\textbf{BfB\_PU\_STU\_Higgs} & \textbf{Precision} & \textbf{Recall} & \textbf{F1-Score} & \textbf{Support} \\
\midrule
0\_0\_0\_0 & 0.9923 & 0.9623 & 0.9771 & 13,584 \\
0\_0\_0\_1 & 0.9693 & 0.9479 & 0.9585 & 5,125 \\
0\_0\_1\_0 & 0.9548 & 0.9684 & 0.9616 & 5,412 \\
0\_0\_1\_1 & 0.9509 & 0.9593 & 0.9551 & 4,100 \\
0\_1\_0\_0 & 0.9726 & 0.9047 & 0.9374 & 3,881 \\
0\_1\_0\_1 & 0.9700 & 0.9199 & 0.9443 & 4,393 \\
0\_1\_1\_0 & 0.9503 & 0.9045 & 0.9268 & 3,549 \\
0\_1\_1\_1 & 0.9448 & 0.9166 & 0.9305 & 5,935 \\
1\_0\_0\_0 & 0.9673 & 0.9539 & 0.9606 & 5,924 \\
1\_0\_0\_1 & 0.9730 & 0.9581 & 0.9655 & 6,970 \\
1\_0\_1\_0 & 0.9595 & 0.9696 & 0.9646 & 5,893 \\
1\_0\_1\_1 & 0.9653 & 0.9858 & 0.9754 & 11,416 \\
1\_1\_0\_0 & 0.9517 & 0.9414 & 0.9465 & 5,444 \\
1\_1\_0\_1 & 0.9395 & 0.9377 & 0.9386 & 5,779 \\
1\_1\_1\_0 & 0.9230 & 0.9463 & 0.9345 & 5,773 \\
1\_1\_1\_1 & 0.9520 & 0.9961 & 0.9735 & 22,962 \\
\bottomrule
\end{tabular}
\caption{Powerset classification metrics for each label combination in the test dataset.}
\label{tab:powerset_clf}
\end{table}

\subsection{Time advantage analysis}
To quantify the practical time advantage of the DL approach, we conducted a timing analysis comparing the traditional physics-based calculations, using \texttt{ScannerS}, with our trained multilabel classifier. The tests were performed on a workstation equipped with an Intel Xeon Silver 4114 CPU (2.20 GHz, 10 cores/20 threads) and 32 GB of RAM. Additionally, to factor out any internal effects on the physics tool, we used the already generated test dataset (116,140 points), and executed the \texttt{check} function of \texttt{ScannerS}, which is designed to work on previously sampled data. We restrict \texttt{ScannerS} to only check (i.e. classify) the four relevant constraints. Hence, eliminating any time taken to generate samples, validate input parameters, or check DM and vacuum stability. We find that \texttt{ScannerS} required 31,644.96 seconds (about 9 hours) to complete the \texttt{check} job. 

As for the DL method, we note that applying the Yeo-Johnson transformation and scaling to the test dataset required 1.06 seconds before running the classifier. During the evaluation of the classifier on the test dataset, we used the default \texttt{Keras} batch size of 32 (not to be confused with batch size used for the training job). This resulted in a prediction time of 103.68 seconds. Then we increased the inference batch size to 2048 and 8192, which resulted in reducing the time to 3.74 and 2.78 seconds, respectively, with no loss in predictive performance. We note that although the workstation included an AMD Radeon Pro WX 5100 GPU, the classifier ran on CPU only, as our standard \texttt{TensorFlow} build could not utilize the GPU. We expect the utilization of GPU to further reduce the time for the classifier to complete its prediction job. Table~\ref{tab:performance_comparison} summarizes the execution times.

\begin{table}[htbp]
\centering
\begin{tabular}{lrr}
\hline
\textbf{Method} & \textbf{Time (seconds)} & \textbf{Speedup Factor} \\
\hline
\texttt{ScannerS} & 31,644.96 & 1.0$\times$ \\
\multicolumn{3}{l}{\emph{Multilabel classifier (CPU only)}} \\
\quad Default (batch size = 32) & 103.68 & 305.2$\times$ \\
\quad Batch size = 2048 & 3.74 & 8,464.9$\times$ \\
\quad Batch size = 8192 & 2.78 & 11,383.4$\times$ \\
\hline
\end{tabular}
\caption{Execution Time Comparison}
\label{tab:performance_comparison}
\end{table}

The results demonstrate a remarkable improvement in computational efficiency. The trained classifier, running on CPU, achieved a speedup factor of over 11,000 compared to the traditional physics-based calculations, while maintaining the high subset accuracy. Indeed, the substantial reduction in computation time, combined with strong performance across different evaluation metrics, suggests that our DL model could serve as an efficient surrogate for rapid parameter space explorations and checks in the DDP-N2HDM, enabling faster theoretical studies and phenomenological analyses.

\subsection{Physics analysis}
\begin{figure}[htp]
    \centering
    \begin{subfigure}[t]{0.48\textwidth}
        \centering
        \includegraphics[width=\textwidth]{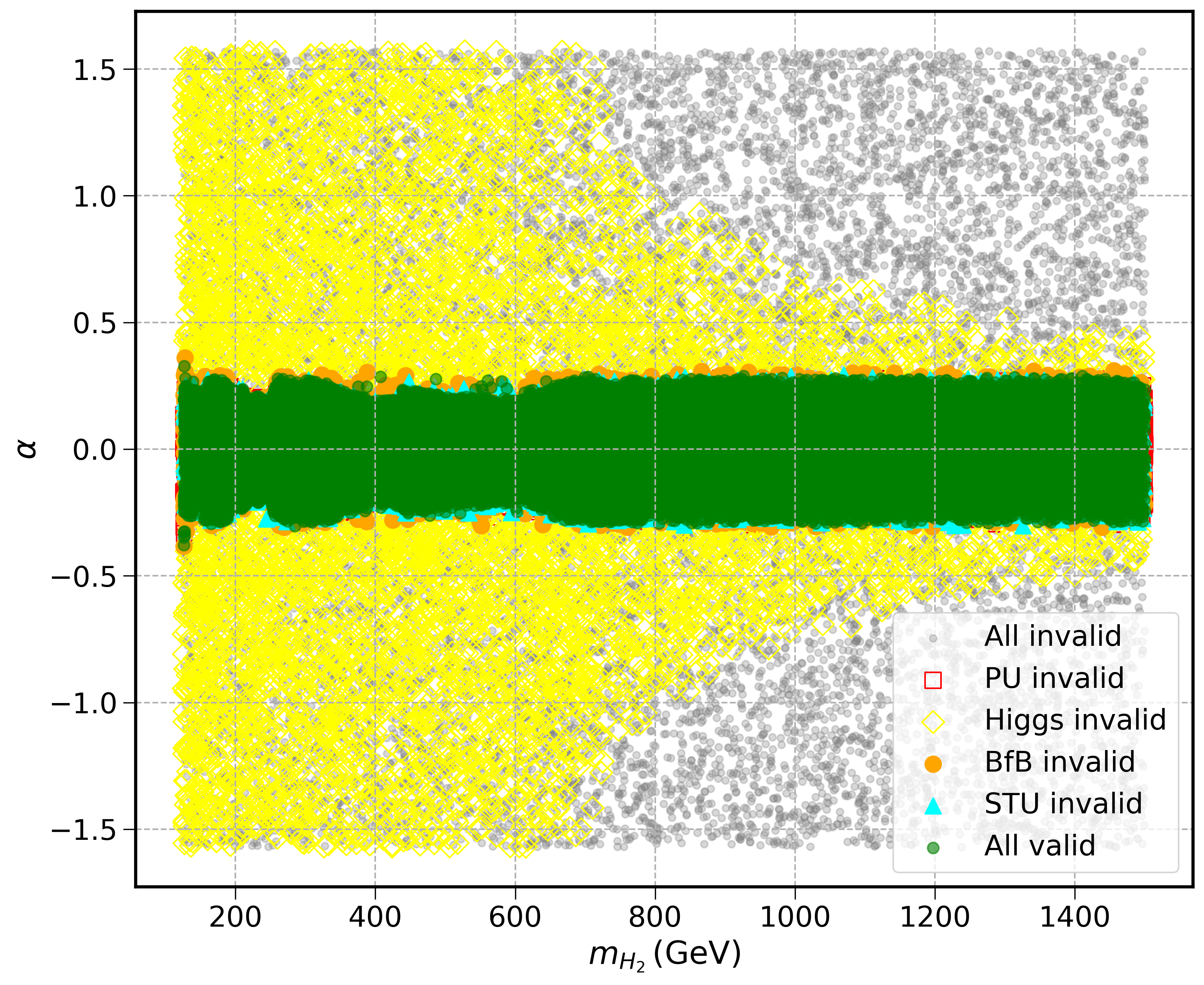}
        \caption{Comparison of all-valid and invalid (BfB, PU, STU, Higgs) scenarios in the $m_{H_2}$ vs. $\alpha$ plane.}
        \label{fig:mH2_alpha}
    \end{subfigure}
    \hfill
    \begin{subfigure}[t]{0.48\textwidth}
        \centering
        \includegraphics[width=\textwidth]{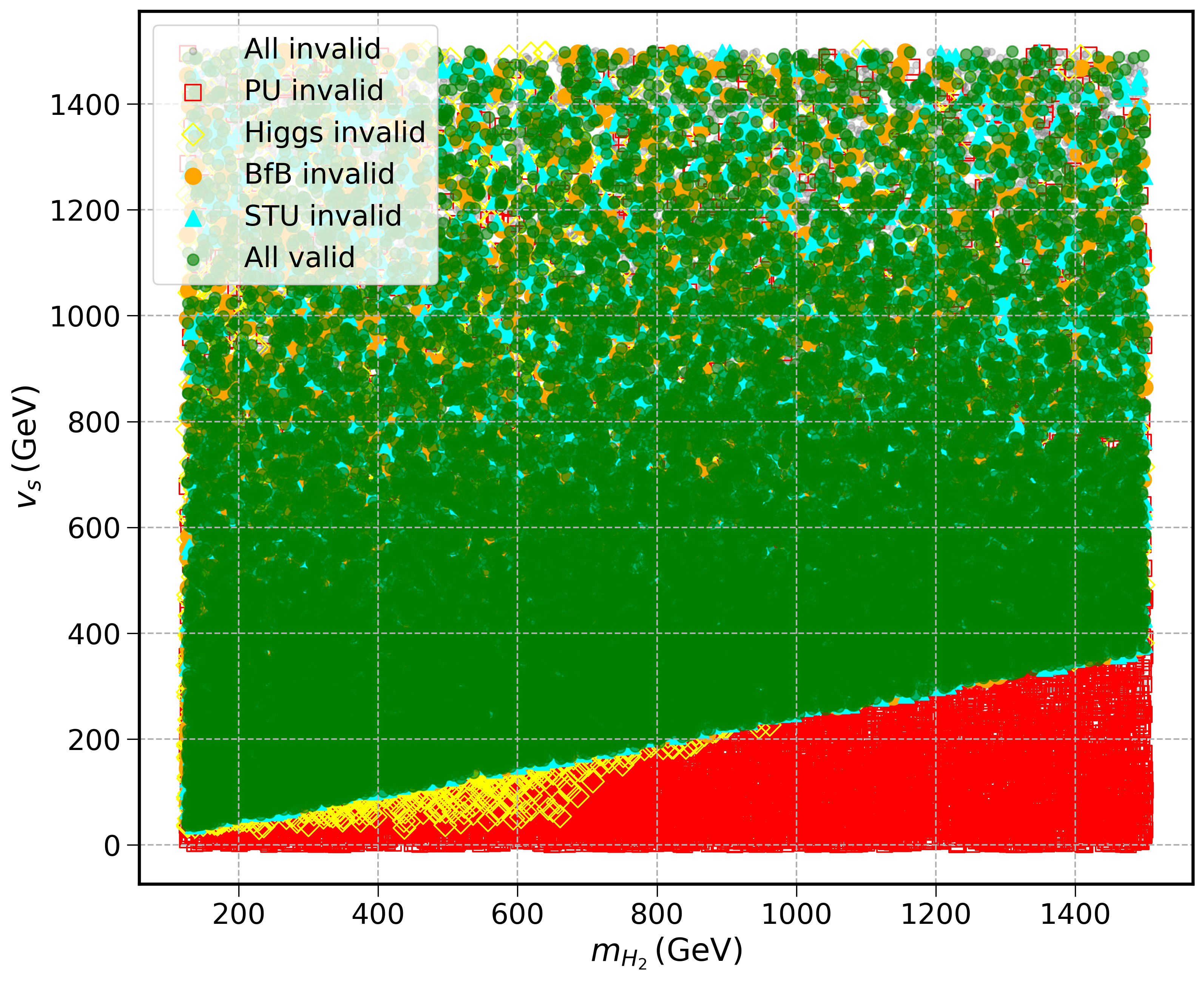}
        \caption{Comparison of all-valid and invalid (BfB, PU, STU, Higgs) scenarios in the $m_{H_2}$ vs. $v_s$ plane.}
        \label{fig:mH2_vs}
    \end{subfigure}
    
    \vspace{1em}
    
    \begin{subfigure}[t]{0.48\textwidth}
        \centering
        \includegraphics[width=\textwidth]{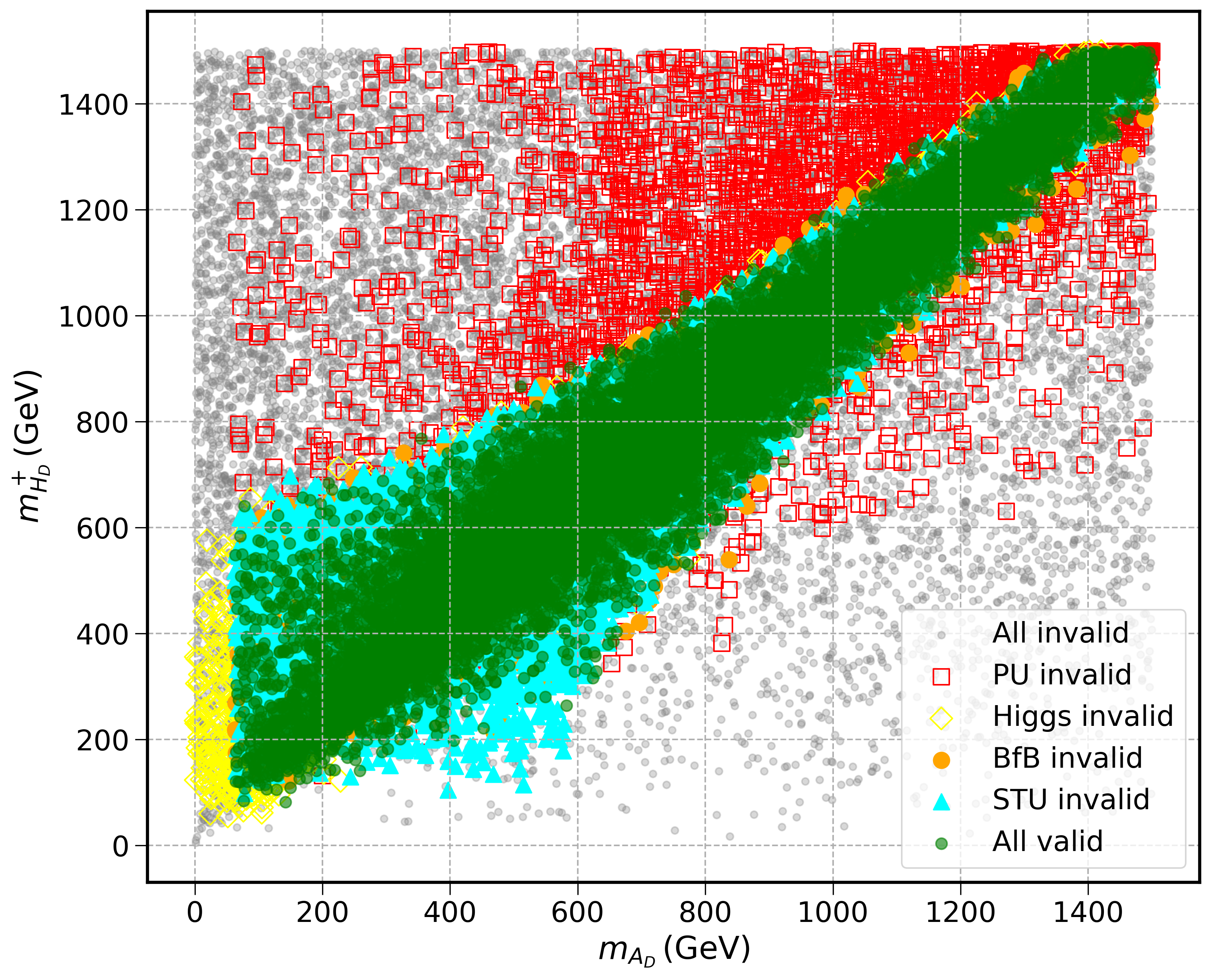}
        \caption{Comparison of all-valid and invalid (BfB, PU, STU, Higgs) scenarios in the $m_A$ vs. $m_{H^\pm}$ plane.}
        \label{fig:mAD_mHDp}
    \end{subfigure}
    \hfill
    \begin{subfigure}[t]{0.48\textwidth}
        \centering
        \includegraphics[width=\textwidth]{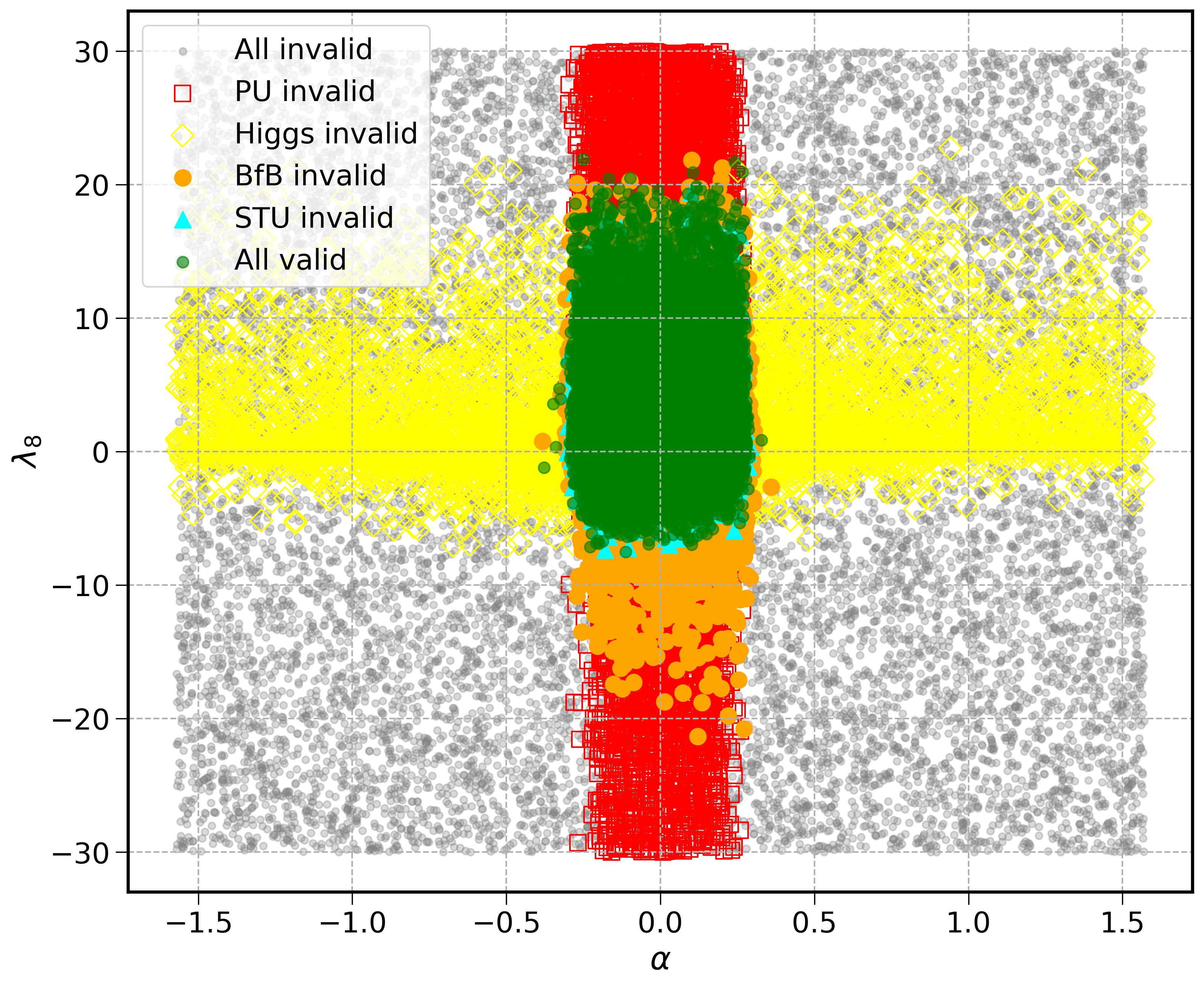}
        \caption{Comparison of all-valid and invalid (BfB, PU, STU, Higgs) scenarios in the $\alpha$ vs. $\lambda_8$ plane.}
        \label{fig:alpha_L8}
    \end{subfigure}
    
    \vspace{1em}
    
    \begin{subfigure}[t]{0.48\textwidth}
        \centering
        \includegraphics[width=\textwidth]{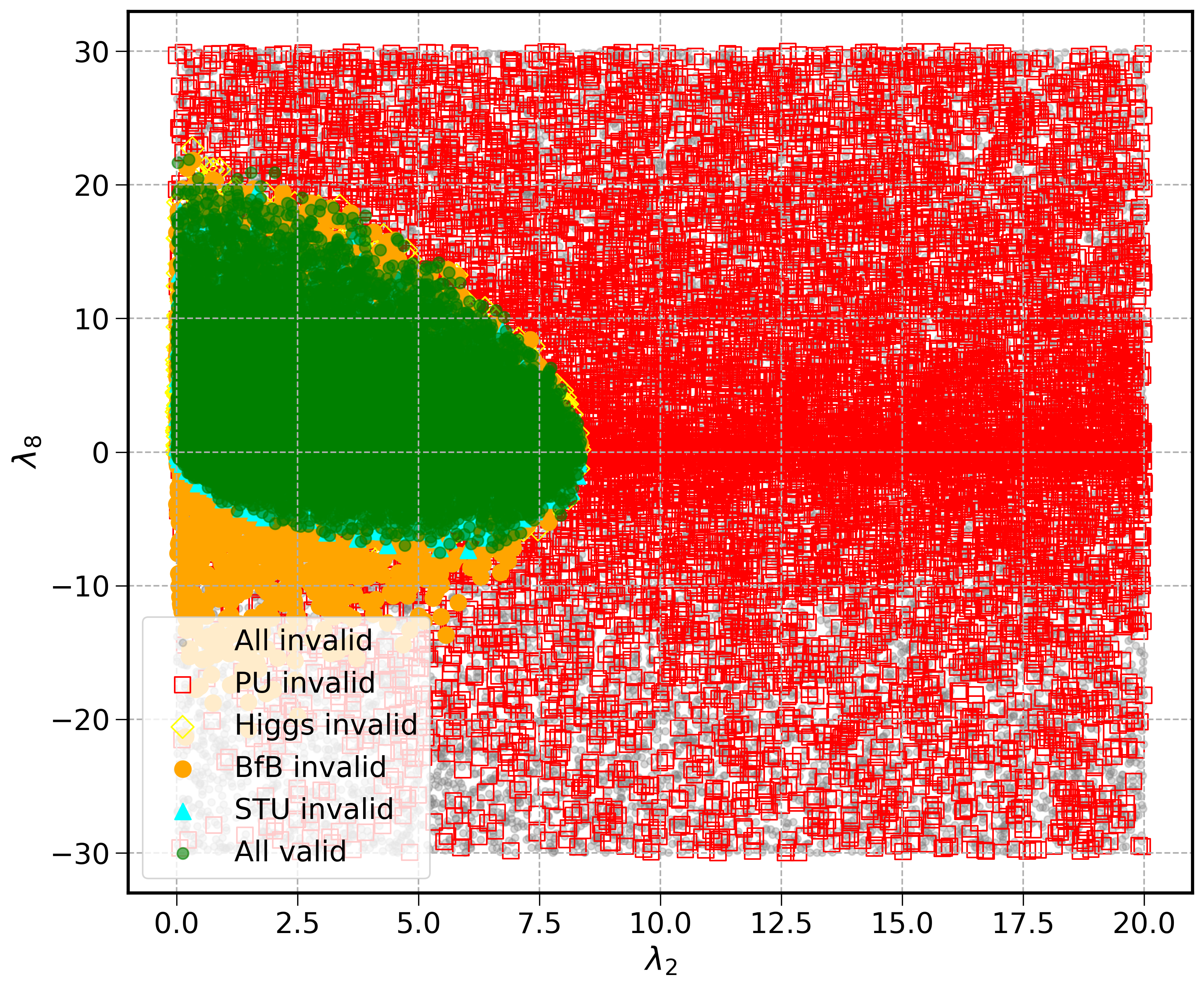}
        \caption{Comparison of all-valid and invalid (BfB, PU, STU, Higgs) scenarios in the $\lambda_2$ vs. $\lambda_8$ plane.}
        \label{fig:L2-L8}
    \end{subfigure}
    \hfill
        \begin{subfigure}[t]{0.48\textwidth}
        \centering
        \includegraphics[width=\textwidth]{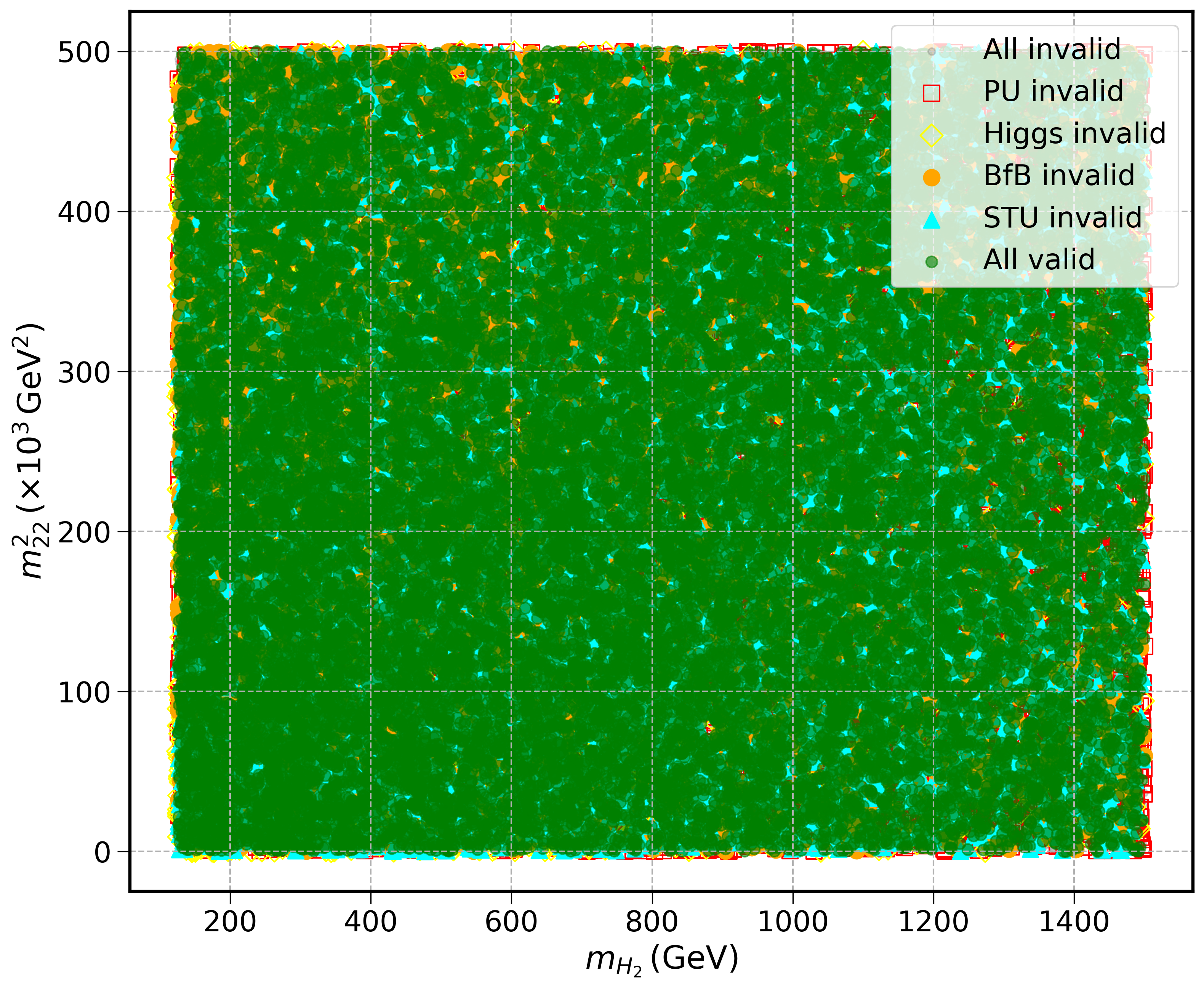}
        \caption{Comparison of all-valid and invalid (BfB, PU, STU, Higgs) scenarios in the $m_{H_2}$ vs. $m_{22}^2$ plane.}
        \label{fig:mh2-m222}
    \end{subfigure}

    \caption{Visualization of the model's parameter space under different constraint validity scenarios. Green points indicate all constraints are satisfied, while red, orange, yellow, and cyan points show where BfB, PU, STU, and Higgs constraints, respectively, are invalid.}
    \label{fig:constraint_comparisons}
\end{figure}

Using the $96\%$ correctly labeled points by the DL model, which comprises 111,504 points, Fig.~\ref{fig:constraint_comparisons} shows the parameter space in the $m_{H_2}-\alpha$, $m_{H_2}-v_s$, $m_{A_D}-m_{H_D}^{+}$, $\alpha-\lambda_8$, and $\lambda_2-\lambda_8$ planes. These were selected after inspecting all possible pairings of the input parameters, as they were found to distinctly show the impact of each constraint.
In each graph, valid points are represented in green, while points that are invalid due to only one of the constraints are shown as follows: red (\textit{PU} invalid), yellow (\textit{Higgs} invalid), orange (\textit{BfB} invalid), and cyan (\textit{STU} invalid). Points that are invalid due to all constraints being violated at the same time are shown in grey. The plotting scheme follows a specific order: the data with all constraints invalid is laid as background, followed by invalid \textit{PU}, invalid \textit{Higgs}, invalid \textit{BfB}, and finally invalid \textit{STU}. Valid points are placed on top of all others, so in regions where valid and invalid points overlap, valid points are prioritized. This ordering reflects the impact of each constraint, with those plotted last having the least impact.

In Fig.~\ref{fig:constraint_comparisons}-a, we observe how \textit{Higgs} constraints rule out the wedge-shaped regions where $m_{H_2}$ is below 1200 GeV, and $|\alpha|>0.3$. In this plane, we can see that this constraint is significant in its own right, while points ruled out due to only one of the other constraints reside beneath the valid region, except for a few points where we see marginal effects from \textit{BfB} and \textit{STU}. The case where all constraints are violated at the same time (grey points) is also significant here as it rules out the top left and right corners. 

Next, Fig.~\ref{fig:constraint_comparisons}-b, shows how \textit{PU} erases the bottom triangle requiring $v_s$ to steadily grow from close to 1 to value above 200 GeV. In this plane, \textit{Higgs} constraints are also apparent in regions around $m_{H_2} \sim 600$ and $ 1 < v_s < 200$ GeV. While the other constraints (\textit{STU} and \textit{BfB}) have a small visible effect for larger values of $v_s$. 

Moreover, Fig.~\ref{fig:constraint_comparisons}-c, illustrates the effect of \textit{STU} in the region where $m_{A_D}$ is between 200 and 600 GeV, and $m_{H_D}^+$ is between 150 and 400 GeV. The \textit{Higgs} constraint rules out regions where $m_{A_D}$ is below 100 GeV, while \textit{PU} rules out distinct regions on the sides of the valid band for values of $m_{H_D}^+ > 600$ GeV. While all constraints being invalid at the same time rule out the remaining parameter space.

The $\alpha$-$\lambda_2$ plane presented in Fig.~\ref{fig:constraint_comparisons}-d clearly illustrates the individual effects of the \textit{Higgs}, \textit{PU}, and \textit{BfB} constraints. While the case of all constraints being invalid excludes the boxed corners in this parameter space.  

Next, the $\lambda_2$-$\lambda_8$ plane in Fig.~\ref{fig:constraint_comparisons}-e again shows the wide impact of \textit{PU} on the parameter space. The effect of \textit{BfB} is also clear, especially on values of $\lambda_2$ between 0 and 7.5, and $\lambda_8$ between 0 and lower -20, which we saw before for $\lambda_8$ in the previously mentioned plot. The \textit{Higgs} and \textit{STU} limits do not play a role in constraining this plane. Fig.~\ref{fig:constraint_comparisons}-f, demonstrates that all four constraints considered in this work do not affect the $m_{H_2}$-$m_{22}^2$ plane, as all of the range of their values are present in the valid points. 

Overall, we see from Figure \ref{fig:constraint_comparisons} that after careful consideration of the main issues surrounding the generation of a multilabel classifier, the trained classifier correctly predicted the four constraints on the parameter space within seconds and with very strong performance. It provided a picture that indeed represents the status of the model, and factual statements about the model and its parameters were made. Therefore, demonstrating the success and promise of this DL method for the task at hand. 

To finalize with a broad picture, we point out only two possible directions that can be considered based on the obtained insights (model status). First, we see that \textit{PU
} constraint had a significant impact, even though the relevant computations within \texttt{ScannerS} implement tree-level precision. However, it was shown in \cite{Goodsell:2018fex} that higher order corrections can affect unitarity constraints, by either closing or opening regions in the parameter space. One can consider this an interesting direction to take for the DDP-N2HDM. 
Second, Higgs constraints also had significant impact on this model, which has a real singlet field. If one extend the model to a complex singlet, then that will lead to an additional CP-odd state $A$, effectively changing the phenomenology and affecting Higgs constraints. For example, if $A$ can decay to $H_2$, then LHC searches for $pp\rightarrow A\rightarrow H_2 Z$ would constrain both $A$ and $H_2$. Also in regions where $H_2$ can decay to $A$, the constraints on $H_2$ from $pp\rightarrow H_2\rightarrow H_1 H_1/VV/f\bar{f}$  would change. Complexifying the singlet will surely have other effects to be determined by a dedicated comparative study. 

\section{Conclusion} \label{discon}
In this work, we have discussed how validating parameter points in BSM extensions involves a complicated chain of checks for several theoretical and experimental constraints. This is usually performed by utilizing dedicated physics tools chained together, requiring a large amount of time. We discussed current advances in utilizing AI methods in the exploration of parameter spaces of SM extensions, and the limitations of single-label tasks in terms of learning the effects of groups of constraints on the models, which is an important aspect of particle phenomenology research. This motivates us to explore, for the first time, the feasibility of training an MLC using DL to perform such a task, focusing on the DDP-N2HDM as a representative multidimensional model. A total of 9 free parameters and 4 constraints (BfB, PU, STU, and Higgs) were taken into account as features and binary target labels. The dataset was generated using a hybrid method combining LHS and random sampling, ensuring good coverage of the parameter space. The dataset was then subjected to appropriate preprocessing, including YJ transformation and scaling.

After experimentation, the DL multilabel classifier was built using relevant metrics for measuring performance. We took every precaution to ensure that it does not overfit the data. Upon applying the resulting classifier to unseen test data, we demonstrated that the DL model performs exceptionally well, achieving a subset accuracy of 0.96, significantly outperforming our baseline RF classifier, which achieved a subset accuracy of 0.78. Additionally, we showed the strong performance of the DL model on individual labels, with all metrics reaching nearly perfect scores. Furthermore, we demonstrated that the classifier effectively learned joint class distributions, capturing the interplay between different constraints, with F1-scores varying between $93\%$ and $98\%$. 

We have carried out a timing analysis, in which we showed that the traditional physics tool required about 9 hours to check the constraints on 116K points, while the trained multilabel classifier performed the job in less than 3 seconds, which is orders of magnitude faster. Hence the classifier can act as a fast surrogate for model status checks with respect to the four considered constraints.  

Using the correctly classified parameter space by the DL model, and with an appropriate choice of parameter planes, we demonstrated the effect of each constraint individually when it is violated, as well as the collective effect when multiple constraints are violated. We observed how the collective violation of constraints rules out certain regions that are not excluded by only one constraint, and pointed out that regions where only one constraint is violated might lead to new research directions. This further demonstrates the success of the method, and its promise for further applications. 

While the DL model was developed specifically for the DDP-N2HDM, the approach is generalizable. To facilitate that, we have provided a python tool \texttt{HEPMLC}, following the methodology described in this paper, to create multilabel classifiers for models implemented in the public tool \texttt{ScannerS} (if the user requires data generation) or for labeled datasets provided by the user.

There are several directions that can be taken in future studies, as the field is still in its infancy. One pressing issue is generating high-quality data in a relatively short time to train multilabel classifiers. Utilizing some of the newly explored single-label methods for fast scans might be an option, by sequentially adapting the fast scanner for each required constraint. This includes utilizing or creating regressors for certain computational jobs. Additionally, it is important to study the impact of new or updated constraints, since the ability to update the classifier smoothly and without requiring full retraining on new large datasets would be a significant advantage. This might be carried out by fine-tuning the classifier with new data; however, the issue of catastrophic forgetting needs to be investigated.    

Finally, as the community is still learning, and work in this direction is progressing, we anticipate many exciting developments that exploit the full power of AI methods in exploring parameter spaces of BSM extensions, with the potential to minimize the need for traditional methods for some time-consuming tasks such as scanning and checking HEP constraints, which are subject to updates due to more precise computations and / or new limits from experimental searches.

\section*{Acknowledgement}
I would like to thank CERN for hospitality during the summer where part of this work was conducted.

\section*{Appendix: Applications with \texttt{HEPMLC}}
\label{sec:appendix_hepmlc}

In this appendix, we describe \texttt{HEPMLC}, a Python-based tool designed for training multilabel classifiers of BSM extensions. The tool is publicly available at \url{https://github.com/drmaien/HEPMLC} and provides a streamlined workflow for training and evaluating neural networks (NNs) to predict theoretical and experimental constraints defined as target (binary) labels.

\subsection*{A.1 Installation and Setup}

\texttt{HEPMLC} is designed to work standalone, if the user provides a dataset containing inputs and constraints (as binary labels). Or to work within \texttt{ScannerS} to generate a dataset for any of the implemented models. In the latter case, after installing \texttt{ScannerS}, \texttt{HEPMLC} can be installed by cloning the repository into the build directory:

\begin{verbatim}
cd ScannerS/build
git clone https://github.com/drmaien/HEPMLC.git
\end{verbatim}


\subsection*{A.2 Tool Structure}

\texttt{HEPMLC} provides a modular structure with the following components:

\begin{itemize}
    \item \texttt{src/preprocessing/}: Data preprocessing and analysis modules.
    \item \texttt{src/modeling/}: NN architecture, optimization, fine-tuning, and training modules.
    \item \texttt{src/utils/}: Utilities for interfacing with ScannerS.
    \item \texttt{notebooks/}: Jupyter notebooks with detailed usage exploiting the full functionality of the tool.
    \item \texttt{HEPMLC.ipynb}: Jupyter notebook with a basic example usage.
\end{itemize}

\subsection*{A.3 Example usage workflow}

A basic example is provided through \texttt{HEPMLC.ipynb}, which guides users through the following steps:

\subsubsection*{A.3.0 Import \texttt{HEPMLC} modules}
Along with essential libraries, we import the tool's modules, which are defined in \texttt{/src} direcotry.

\begin{lstlisting}
# Import required modules. These are defined in /src.
from utils.model_reader import ModelReader
from utils.scanner_runner import ScannerRunner
from preprocessing.preprocessor import FeaturePreprocessor
from modeling.architecture import ModelBuilder
from modeling.trainer import ModelTrainer
\end{lstlisting}

\subsubsection*{A.3.1 Physics model selection}
Users can select and configure any physics model available in ScannerS. For each model, the tool displays:
\begin{itemize}
    \item Available input parameters and their ranges.
    \item Theoretical and experimental constraints.
    \item Configuration options for constraints (apply/ignore/skip).
\end{itemize}
\begin{lstlisting}
# List available models
model_reader = ModelReader(scanner_path=scanner_path)
print("Available Models:")    
\end{lstlisting}
Output:
\begin{verbatim}
Available Models:
--------------------------------------------------
Complex 2HDM Flipped (C2HDM_FL.ini)
Complex 2HDM Lepton Specific (C2HDM_LS.ini)
Complex 2HDM Type 1 (C2HDM_T1.ini)
Complex 2HDM Type 2 (C2HDM_T2.ini)
CP-Violating Dark Matter (CPVDM.ini)
Complex Singlet Broken Phase (CxSMBroken.ini)
Complex Singlet Dark (CxSMDark.ini)
N2HDM Broken Type 2 (N2HDMBroken_T2.ini)
N2HDM Dark D (N2HDMDarkD.ini)
N2HDM Dark SD (N2HDMDarkSD.ini)
N2HDM Dark S Type 1 (N2HDMDarkS_T1.ini)
Real 2HDM Flipped (R2HDM_FL.ini)
Real 2HDM Lepton Specific (R2HDM_LS.ini)
Real 2HDM Type 1 (R2HDM_T1.ini)
Real 2HDM Type 2 (R2HDM_T2.ini)
TRSM Broken Phase (TRSMBroken.ini)  
\end{verbatim}

\begin{lstlisting}
# Select model and configure
selected_model = "N2HDMDarkD.ini"  # Change this to your chosen model
features, constraints = model_reader.read_model(selected_model)    
\end{lstlisting}
Output:
\begin{verbatim}
Features and ranges:
- mHa: [125.09, 125.09]
- mHb: [50.0, 1000.0]
- mHD: [1.0, 1500.0]
- mAD: [1.0, 1500.0]
- mHDp: [1.0, 1500.0]
- alpha: [-1.57, 1.57]
- m22sq: [0.001, 500000.0]
- L2: [0.0, 20.0]
- L8: [-30.0, 30.0]
- vs: [1.0, 1500.0]

Constraints:
- BfB: ignore
- Uni: ignore
- STU: ignore
- Higgs: ignore
- VacStab: skip
- DM: skip
\end{verbatim}

\subsubsection*{A.3.2 Data generation}
The tool interfaces with ScannerS to:
\begin{itemize}
    \item Generate training data with specified parameters (features and labels).
    \item Analyze feature distributions, and class distributions of the target labels.
    \item Provide recommendations for additional scans if severe class imbalance is detected.
\end{itemize}

\begin{lstlisting}
# Configure scan parameters
n_points = 1000  # Number of points to generate
output_file = os.path.join(build_dir, 'HEPMLC', results_dir, "scan_data.tsv")

# Run scan
scanner = ScannerRunner()
scanner.run_scan(selected_model, n_points, output_file)

# Analyze class balance
feature_cols = ['mH2', 'mHD', 'mAD', 'mHDp', 'alpha', 'L2', 'L8', 'vs', 'm22sq']
label_cols = ['valid_BFB', 'valid_Uni', 'valid_STU', 'valid_Higgs']

stats = scanner.analyze_class_balance(output_file, label_cols)
scanner.plot_class_distribution(stats, os.path.join(results_dir, 'class_distribution'))    
\end{lstlisting}
Output:
\begin{figure}[htb]
    \begin{jupyteroutput}
        \includegraphics[width=0.8\textwidth]{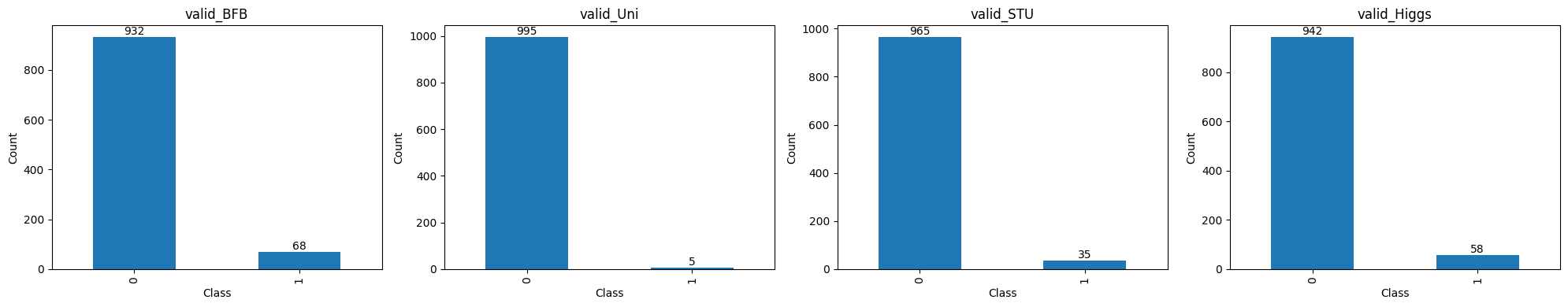}
    \end{jupyteroutput}
    \caption{Output showing class balance for each label.}
    \label{fig:ex1}
\end{figure}

\begin{figure}[htb]
    \begin{jupyteroutput}
        \includegraphics[width=0.8\textwidth]{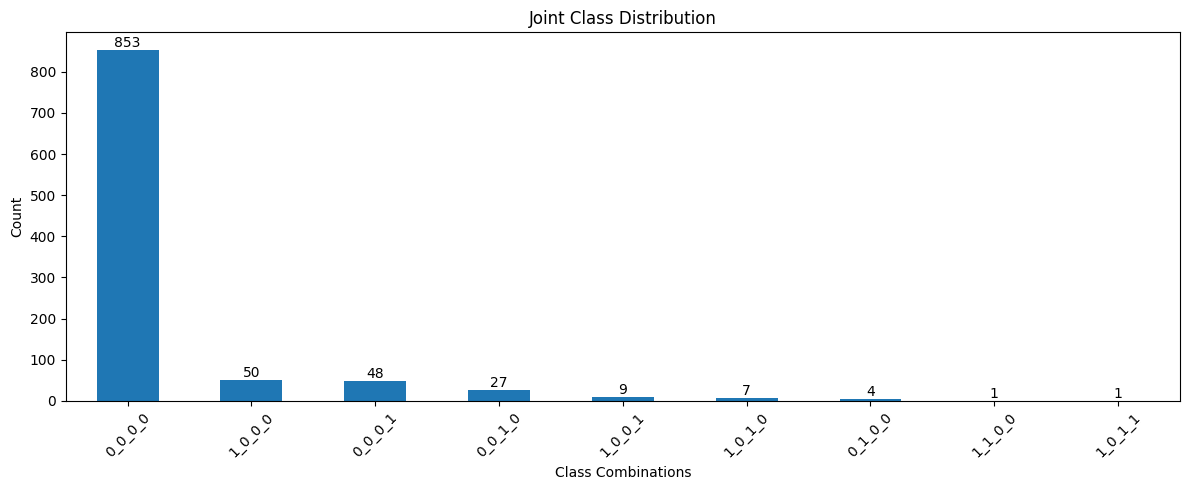}
    \end{jupyteroutput}
    \caption{Output showing joint class distributions.}
    \label{fig:ex2}
\end{figure}

\subsubsection*{A.3.4 Data preprocessing and spliting}

\begin{lstlisting}
# Configure preprocessing
apply_yj = True      # Apply Yeo-Johnson transformation
apply_scaler = True  # Apply Standard scaling

# Initialize preprocessor
preprocessor = FeaturePreprocessor(apply_yj=apply_yj, apply_scaler=apply_scaler)

# Split data (70-15-15)
X = data[feature_cols]
y = data[label_cols]

X_temp, X_test, y_temp, y_test = train_test_split(X, y, test_size=0.15, random_state=42)
X_train, X_val, y_train, y_val = train_test_split(X_temp, y_temp, test_size=0.176, random_state=42)

print(f"Training set: {len(X_train)} samples")
print(f"Validation set: {len(X_val)} samples")
print(f"Test set: {len(X_test)} samples")

# Preprocess data
X_train_processed = preprocessor.fit_transform(X_train)
X_val_processed = preprocessor.transform(X_val)
X_test_processed = preprocessor.transform(X_test)

# Save preprocessor
preprocessor.save_transformers(os.path.join(results_dir, 'preprocessor'))
\end{lstlisting}

\subsubsection*{A.3.5 Model configuration and training}
Users can modify the model's (hyper)parameters through the interface. Training progress is monitored with multiple metrics:
\begin{itemize}
    \item Subset accuracy
    \item Hamming loss
    \item Matthews correlation coefficient
    \item Macro F1 score
\end{itemize}

\begin{lstlisting}
# Model configuration (default parameters from paper)
model_params = {
    'n_layers': 2,
    'n_units_0': 64,
    'n_units_1': 128,
    'activation': 'relu',
    'dropout_rate': 0.01, #0.117,
    'apply_batch_norm': True,
    'optimizer': 'adam',
    'regularization': None,
    'reg_lambda': 0.05,
    'learning_rate': 0.000263,
    'batch_size': 64
}

# Training configuration
training_params = {
    'epochs': 15,
    'patience': 5,  # Early stopping patience
}

# Initialize model and trainer
builder = ModelBuilder(
    input_shape=(len(feature_cols),),
    num_outputs=len(label_cols)
)

trainer = ModelTrainer(
    model_builder=builder,
    feature_cols=feature_cols,
    label_cols=label_cols,
    output_dir=os.path.join(results_dir, 'model')
)

# Train model
model = trainer.train(
    X_train=X_train_processed,
    y_train=y_train,
    X_val=X_val_processed,
    y_val=y_val,
    params=model_params,
    epochs=training_params['epochs']
)
\end{lstlisting}

\subsubsection*{A.3.6 Evaluation}
The tool provides comprehensive evaluation including:
\begin{itemize}
    \item Individual label performance metrics.
    \item Confusion matrices.
    \item Powerset-based evaluation.
    \item Joint label distribution analysis.
\end{itemize}

\begin{lstlisting}
# Evaluate on test set
trainer.evaluate(
    model=model,
    X_test=X_test_processed,
    y_test=y_test
)

print("\nAll results have been saved in the Results directory.")
\end{lstlisting}

The output of the last two steps contains detailed \texttt{csv} files and plots on the training history, and performance of the classifier. Example outputs can be found in the Results folder within the Github repository.

%

\end{document}